\documentclass[aps,pra,twocolumn,superscriptaddress, nofootinbib]{revtex4-1}

\usepackage{amsmath}
\usepackage{amssymb}
\usepackage{mathrsfs}
\usepackage{verbatim}
\usepackage{epsfig}
\usepackage{color}
\usepackage{soul}
\usepackage[normalem]{ulem}
\usepackage[bookmarks]{hyperref}
\usepackage{orcidlink}
\usepackage{graphicx}
\usepackage{derivative}
\usepackage{aligned-overset}
\usepackage{nicefrac}
\usepackage{mhchem}
\usepackage{relsize}
\usepackage[normalem]{ulem}
\usepackage{enumerate}
\usepackage{chemformula}

\graphicspath{{.}{./EPS/}}

\newcommand* {\bra}[1]{\ensuremath{\langle {#1} |}}
\newcommand* {\ket}[1]{\ensuremath{| {#1} \rangle}}
\newcommand* {\Tr}[0]{\ensuremath{\text{Tr} }}
\newcommand* {\Var}[0]{\ensuremath{\text{Var} }}

\DeclareMathOperator{\sinc}{sinc}

\begin{document}

\title{Bayesian and frequentist estimators for the transition frequency of a driven two-level quantum system}

\author{Chun Kit Dennis Law}
\affiliation{Forschungszentrum Jülich, Institute of Quantum Control,
Peter Grünberg Institut (PGI-8), 52425 Jülich, Germany}

\author{J\'ozsef Zsolt Bern\'ad}
\email{j.bernad@fz-juelich.de}
\affiliation{Forschungszentrum Jülich, Institute of Quantum Control,
Peter Grünberg Institut (PGI-8), 52425 Jülich, Germany}
\affiliation{HUN-REN Wigner Research Centre for Physics, Budapest, Hungary}

\begin{abstract}
The formalism of quantum estimation theory with a specific focus on classical data postprocessing is applied to a two-level system driven by an external gyrating magnetic field. We employed both Bayesian and frequentist approaches to estimate the unknown transition frequency. In the frequentist approach, we have shown that only reducing the distance between the classical and the quantum Fisher information does not necessarily mean that the estimators as functions of the data deliver an estimate with desirable accuracy, as the classical Fisher information takes small values. We have proposed and investigated a cost function to account for the maximization of the classical Fisher information and the minimization of the aforementioned distance. Due to the nonlinearity of the probability mass function of the data on the transition frequency, the minimum variance unbiased estimator may not exist. The maximum likelihood and the maximum {\it a posteriori} estimators often result in ambiguous estimates, which in certain cases can be made unambiguous upon changing the parameters of the external field. It is demonstrated that the minimum mean-square error estimator of the Bayesian statistics provides unambiguous estimates. In the Bayesian approach, we have also investigated the effects of noninformative and informative priors on the Bayesian estimates, including a uniform prior, Jeffrey's prior, and a Gaussian prior.    
\end{abstract}

\maketitle

\section{Introduction}
\label{sec:Intro}

Parameter estimation is a central topic in statistics and plays a pivotal role in science. It is employed when one makes indirect observations of a parameter of interest and then tries to infer its value
with sufficiently high accuracy. Usually, this is done for understanding the time evolution of a dynamical system or designing technological devices such as digital communication systems \cite{vanTrees}. This is an inverse problem and common challenges include the sensitivity to noise and modeling inaccuracies. Thanks to the recent developments in quantum technology, specifically in the field of quantum metrology, one can achieve estimation accuracy of physical parameters, which surpasses the classical limits with the use of entangled states \cite{PezzeRev, Kok}. Quantum metrology has seen advances in single-parameter estimation such as the detection of gravitational waves \cite{Ganapathy}, technological progresses in atomic clocks \cite{Pedrozo, Robinson}, programmable quantum sensors \cite{Marciniak}, and quantum thermometry \cite{Rubio_2020, Rubio_2022}. This subfield of quantum technology has also been extended to cases, in which simultaneous estimations of multiple parameters are performed \cite{Demkowicz}.

In this work, we focus on the classical postprocessing of the data collected from a two-level quantum system (TLS). In particular, we are interested in the properties of both Bayesian and frequentist estimators of the transition frequency, when the TLS is subject to an external time-dependent drive. Most inference investigations of quantum systems tend to emphasize the saturation of the Cram\'er-Rao bound involving the classical Fisher information (CFI), which determines the accuracy of the estimates, and its upper bound, the quantum Fisher information (QFI) \cite{Haase, Toscano, Yu}. These quantities are important benchmarks against which we can compare the performance of the estimators. However, they do not carry any information about the construction of the estimators as functions of the dataset. The determination of the optimal or at least well-performing estimators is central to our study presented here. In general, a quantum estimation problem consists of searching for optimal positive operator measures (POVM) and classical estimators \cite{Hayashi}. However, the choice of an efficient classical estimator is often not obvious, when the data are characterized by a probability distribution or mass function pertinent to a quantum system. The problem can be described operationally as a nonlinear dependence of the probability function on the parameter to be estimated (see Sec. $2.4$ in Part I of Ref.\cite{vanTrees}). In this context, the role of the external drive will be the modulation of the TLS in such a way that the investigated estimators perform better.

We explore this scenario for a prototype model studied by Rabi in $1937$ \cite{Rabi}. Here, a time-dependent magnetic field
drives the TLS and enables transitions between the two energy states. The emitted photons are collected by a photodetector, which produces binary output data. This model has an analytical solution, which is essential for the construction of the estimators. On the other hand, the model has a long history, see for example, an older review \cite{Donin} or recent developments \cite{Dittrich, Stievater, Pedernales}. The literature of this model is considerable, and therefore the study of the associated estimation strategies may prove to be useful to the fields of research in which the model is employed. Thus, obtaining an accurate estimate of the transition frequency of this particular TLS can be crucial for the development of future quantum devices. 

\begin{figure*}[htbp]
    \begin{center}
    \includegraphics[width=0.9\textwidth]{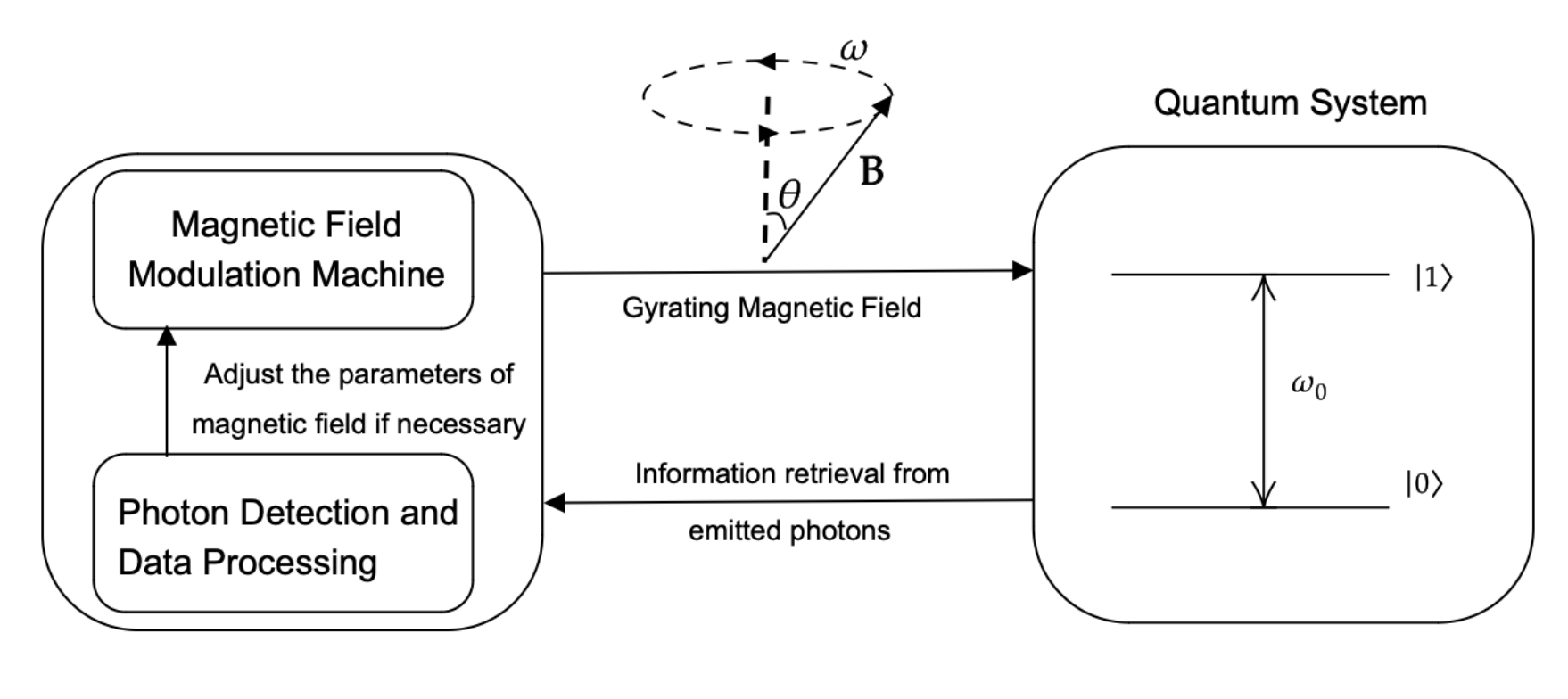}
    \end{center}
    \caption{Schematic representation of a quantum estimation scenario based on a two-level system (TLS) driven by a circularly polarized magnetic field. The magnetic-field modulation machine generates the rotating magnetic field $\mathbf{B}$ at an angular speed $\omega$ with an angle $\theta \in (0,\pi)$ about the axis of gyration. The photon detector registers the emitted photons from the TLS and then the data are processed. The parameters of the magnetic field are adjusted accordingly to achieve optimal estimation of the transition frequency $\omega_0$.}
    \label{fig:Control}
\end{figure*}

Finally, we are also interested in the difference between the estimators offered by the Bayesian and frequentist approaches.
The Bayesian estimation requires the incorporation of the existing knowledge before the measurements are done and this means, operationally, that a prior probability distribution function is employed, i.e., the parameter subject to estimation is treated as a random variable. Although the two approaches of thoughts have different interpretations of the probability \cite{Fine}, the current article places the emphasis only on the methodological aspects. In particular, we set out to investigate the choice of informative and noninformative priors \cite{BernardoBook} which are usually not the focus of previous studies based on Bayesian statistics \cite{Teklu_2010}, but may have unforeseen impacts on the parameter estimation. 

This article is organized as follows. In Sec. \ref{2}, we briefly discuss the Rabi model, where the solutions are given. In Sec. \ref{3}, a condensed overview of quantum estimation theory and classical data postprocessing is presented, where we provide a summary of different frequentist and Bayesian estimators. The probability mass function for the photon measurements and the underlying assumption are described. The various priors deployed for the Bayesian estimators are also explained. In Sec. \ref{4}, the numerical results are presented and the obtained estimates for the transition frequency of the TLS are discussed in depth. In Sec. \ref{5} we summarize and draw our conclusions. Some details supporting the main text are collected in three appendices.

\section{Model}
\label{2}

 We consider a two-level system (TLS), which is driven by a circularly polarized magnetic field. The analytical solution to this model is known \cite{Rabi}. The two states of the TLS are the ground state $\ket{0}=(0,1)^T$ and the excited state $\ket{1}=(1,0)^T$ ($T$ indicates the transpose of the vector). The free Hamiltonian of the TLS reads ($\hbar=1$)
\begin{equation}
    H_{0} = \frac{1}{2} \omega_{0} \sigma_{z},
\end{equation}
where $\sigma_{z}=\ket{1}\bra{1}-\ket{0}\bra{0}$, i.e., a Pauli matrix, and $\omega_{0}$ is the transition frequency. The TLS is placed in a circularly polarized magnetic field $\mathbf{B}$ as shown in Fig. \ref{fig:Control}. The magnetic field has a constant magnitude $B_{0}$ and is rotating at an angular speed $\omega$ at an angle $\theta$ about the axis of gyration. This magnetic field at time $t$ is parametrized as $\mathbf{B}(t) = (B_{0} \sin \theta \cos (\omega t), B_{0} \sin \theta \sin (\omega t), B_{0} \cos \theta)^{T}$ with $\theta \in (0,\pi)$. In terms of $\sigma_{z}$ and the other Pauli matrices $\sigma_{x} = \ket{0}\bra{1}+\ket{1}\bra{0}$ and $\sigma_{y} = i\ket{0}\bra{1}-i\ket{1}\bra{0}$, the interaction between the TLS and the magnetic field is described by 
\begin{equation}
    H_{I}(t) = \gamma \mathbf{B}(t) \cdot \boldsymbol\sigma,
\end{equation}
with $\boldsymbol\sigma = (\sigma^{x}, \sigma^{y}, \sigma^{z})^{T}$. The coupling strength of the magnetic field to the TLS is determined by the factor $\gamma=g\mu_{B}/2$, where $g$ and $\mu_{B}$ are the Land\'{e} $g$-factor and the Bohr magneton, respectively. 

The time-dependent quantum state of the system is
\begin{equation}
\ket{\Psi(t)}= c_0(t) \ket{0} + c_1(t) \ket{1}  
\end{equation}
where the probability amplitudes $c_{0}(t)$ and $c_{1}(t)$ satisfy $|c_0(t)|^2+|c_1(t)|^2=1$. The solution to the Sch\"{o}dinger equation governed by $H_{0}+H_{I}(t)$ was obtained by Rabi for arbitrary initial conditions \cite{Rabi}, see the details in Appendix  \ref{AppendixA}.

Although the populations of the two states are oscillating, we want to make sure that the photons are emitted starting from $t=0$. Therefore, we are interested only in a particular situation, when $c_1(0)=1$ and $c_0(0)=0$. In this prototype model, the probability $|c_0(t)|^2$ is associated with the photon detection as long as the whole measurement setup is considered ideal. The solution to the Sch\"{o}dinger equation for this initial condition reads
\begin{equation}\label{C0}
    c_{0}(t) = -2i e^{-\frac{i\omega t}{2}} \frac{\gamma B_{0} \sin \theta}{q} \sin \left( \frac{qt}{2} \right)
\end{equation}
and
\begin{align}\label{C1}
    c_{1}(t) &= e^{\frac{i\omega t}{2}} \left[ \cos \left(\frac{qt}{2} \right) \right. \notag \\
    &\left. \qquad - i  \frac{\omega-\omega_{0} - 2 \gamma B_{0} \cos \theta}{q} \sin \left(\frac{qt}{2} \right) \right] ,
\end{align}
where 
\begin{equation}\label{q}
    q=\sqrt{(\omega_{0}-\omega)^{2} + 4 \gamma B_{0} \cos \theta (\omega_{0}-\omega) + 4 \gamma^{2} B_{0}^{2}} . 
\end{equation}

These equations yield a complete description of our system and we employ them in our subsequent investigation for the optimal estimation of the unknown parameter $\omega_0$. Fig. \ref{fig:Control} shows the schematic of the estimation process. We use the rotating magnetic field as a modulation machine, which enables transitions between the excited state and the ground state. Thus, the emission of photons is regulated by the time-dependent magnetic field. In our model, the field is treated classically. Still, the information concerning the TLS is then retrieved by detecting the emitted photons, which is characterized by the probability distribution of the ground state. To obtain efficient estimates of $\omega_{0}$, the angular speed $\omega$, the field strength $B_{0}$, and the angle $\theta$ about the axis of gyration will be altered. 

The detection process requires a predefined time $t$, which has to be shorter than the operating gate window of the single photon detector \cite{Marsili, Zhao}. Then, the detector receives a photon with probability $|c_0(t)|^2$ and after that, the TLS is reinitialized to the excited state, which is accompanied by resetting the detector. This allows us to describe subsequent observations as mutually independent.
In this sense, the obtained data are a realization of independent and identically distributed (i.i.d.) random variables that take the value $1$, i.e., one photon is detected, and the value $0$, when no detection occurs. The detectors are considered ideal, because our focus lies on the methodology concerning both frequentist and Bayesian approaches in this particular estimation scenario.

\section{Quantum Estimation Theory}
\label{3}

In this section, we provide concepts about parameter estimation in quantum theory, which is extensively discussed in the literature \cite{Helstrombook, Paris, Holevo, Hayashi, Demkowicz}. Our focus lies on
the classical estimators, but we discuss also the symmetric logarithmic derivative, quantum Cram\'er–Rao bound and the related quantum Fisher information (QFI). In particular, we briefly review the derivation of QFI to investigate the estimation of the transition frequency $\omega_{0}$. Since both frequentist and Bayesian approaches are deployed in estimating $\omega_{0}$, this section commences with the frequentist version of the classical and quantum Fisher information. Then we conclude with a discussion of the Bayesian counterparts. 

\subsection{Cram\'er-Rao inequalities}

{\it Frequentist approach.} In this case, $\omega_0$ is considered to be a nonrandom parameter. Our starting point is the experimentally generated $N$-point dataset ${\bf x}=\{x_1,x_2, \dots, x_N\}$ with $x_i \in \{0,1\}$. As we argued before, this data is realized after i.i.d. observations. The measurement result $x = 1$ with probability $p(1;\omega_{0}) = \vert c_{0}(t) \vert^{2}$ 
($x=0$ with probability $p(0;\omega_{0}) = \vert c_{1}(t) \vert^{2}$) indicates that the system is in the ground state (excited stated) and a photon is (not) detected. The joint probability mass function (PMF) of the data is a function dependent on $\omega_{0}$, whose expression is given by
\begin{equation}
    p({\bf x}; \omega_0)= \binom{N}{k} \prod^N_{i=1} p^{x_i}(1; \omega_0)\left[1-p(1; \omega_0)\right]^{1-x_i},
    \label{eq:datapmf}
\end{equation}
where $k$ is the number of photon counts and
\begin{equation}
    \binom{N}{k} =\frac{N!}{k!(N-k)!}
\end{equation}
is the binomial coefficient. For example, the above PMF treats equally the following detection outputs for $N=3$ and $k=2$: $\{1,1,0\}$, $\{1,0,1\}$, and $\{0,1,1\}$.

We wish to determine the point estimator $\hat{\omega}_0({\bf x})$,  which is a function of the dataset. The classical Fisher Information (CFI) defines both a lower bound for the mean-square error and the width of the Gaussian probability distribution function of the maximum likelihood estimator for asymptotically large data size \cite{Kay, vanTrees}. Now, provided that $p({\bf x}; \omega_0)$ satisfies the regularity condition, which we discuss in Appendix \ref{AppendixB}, the CFI for one sample $x$ reads
\begin{equation}
\label{eq:likelihood,omega}
 \mathcal{F}_{\omega_{0}}^{\text{freq}} = E\left[\left(\frac{\partial \ln p(x;\omega_{0})}{\partial \omega_{0}}\right)^2\right],   
\end{equation}
where $E[\cdot]$ denotes the expectation value with respect to the $p(x;\omega_{0})$. In the case of an unbiased estimator, the mean-square error is equal to the variance of the estimator and the Cram\'er-Rao inequality reads \cite{Kay}
\begin{equation}\label{CRB}
    \Var\left[\hat{\omega}_{0}({\bf x})\right] = E\left[\hat{\omega}_{0}^{2}({\bf x}) \right]-E[\hat{\omega}_{0}({\bf x})]^{2} \geq \frac{1}{N \mathcal{F}_{\omega_{0}}^{\text{freq}}}.
\end{equation}

The goal of general estimation problems in quantum systems is to find the efficient positive operator-valued measure (POVM) which optimizes the classical data processing \cite{Hayashi}.  A POVM is a set of positive-semidefinite matrices ${\bf M}=\{M_{x}\}_{x\in J}$, where $J$ is the set of measurement outcomes, satisfying  $\sum_{x \in J} M_{x} = I$, where $I$ is the identity matrix. In this study, we are only concerned with projective measurements, i.e., a special POVM, whose measurement operators are $M_{0}=\ket{0}\bra{0}$ and $M_{1}= I_2 - \ket{0}\bra{0} = \ket{1}\bra{1}$, where $I_2$ is the $2\times 2$ identity matrix. In fact, to every measurement scenario there exists a CFI, which we denote by $\mathcal{F}_{\omega_{0}}^{\text{freq}}({\bf M})$. Now, maximizing the CFI over all all POVMs \cite{Paris}, we obtain the quantum Fisher Information (QFI): 
\begin{equation}
    \mathcal{H}_{\omega_{0}}^{\text{freq}} = \max_{{\bf M}} \mathcal{F}_{\omega_{0}}^{\text{freq}}({\bf M}). 
\end{equation} 
Thus, the CFI is bounded above by the QFI, i.e., $\mathcal{F}_{\omega_{0}}^{\text{freq}}\leq \mathcal{H}_{\omega_{0}}^{\text{freq}}$. An alternative expression for the QFI reads $\mathcal{H}_{\omega_{0}}^{\text{freq}}= \Tr\left[\rho L_{\omega_{0}}^{2}\right]$, where the symmetric logarithmic derivative (SLD) $L_{\omega_{0}}$ satisfies the relation
\begin{equation}\label{SLD}
    \frac{\partial \rho}{\partial \omega_{0}} = \frac{1}{2} (L_{\omega_{0}} \rho + \rho L_{\omega_{0}})
\end{equation}
in terms of the density matrix. In our case, the density matrix reads
\begin{equation}\label{DenMat}
    \rho = \ket{\Psi(t)}\bra{\Psi(t)} = 
    \begin{pmatrix}
        \rho_{00} & \rho_{01} \\
        \rho_{01}^{*} & 1-\rho_{00} 
    \end{pmatrix}
\end{equation}
whose independent components are $\rho_{00}=\vert c_{0}(t)\vert^{2}$ and $\rho_{01}=c_{0}(t) c_{1}^{*}(t)$. Then, the CFI and the QFI are given by the expressions (see Appendix \ref{AppendixC}) 
\begin{equation}\label{CFI}
    \mathcal{F}_{\omega_{0}}^{\text{freq}}= \frac{1}{\rho_{00}(1-\rho_{00})} \left[\frac{\partial \rho_{00}}{\partial \omega_{0}} \right]^{2}
\end{equation}
and
\begin{equation}\label{QFI}
    \mathcal{H}_{\omega_{0}}^{\text{freq}}= \Tr\left[\rho L_{\omega_{0}}^{2}\right]=4 \left[ \left(\frac{\partial \rho_{00}}{\partial \omega_{0}} \right)^{2} + \left \vert \frac{\partial \rho_{01}}{\partial \omega_{0}} \right \vert^{2} \right],
\end{equation}
respectively. Based on the results in Sec. \ref{2}, we have
\begin{widetext}
    \begin{equation}\label{TLSCFI}
        \mathcal{F}_{\omega_{0}}^{\text{freq}} = \frac{16\gamma^{2}B_{0}^{2} \sin^{2} \theta}{q^{4}} \left( \omega - \omega_{0} - 2 \gamma B_{0} \cos \theta \right)^{2} \frac{\left[ \sin \left( \frac{qt}{2} \right) - \frac{qt}{2} \cos \left( \frac{qt}{2} \right)  \right]^{2}}{q^2 - 4 \gamma^2 B^2_{0} \sin^2 \theta \sin^{2} \left( \frac{qt}{2} \right)}, 
    \end{equation}   
and
    \begin{align}\label{TLSQFI}
        \mathcal{H}_{\omega_{0}}^{\text{freq}} = \frac{4\gamma^{2}B_{0}^{2}\sin^{2} \theta}{q^{6}} &\left\{ \frac{4\gamma^{2} B_{0}^{2}\sin^{2} \theta}{q^{2}} \left(\omega-\omega_{0}- 2 \gamma B_{0} \cos \theta\right)^2 \left[  2-2\cos(qt) - qt \sin (qt)  \right]^{2} \right. \notag\\
        & \qquad \left. + \left[ \left( q^2- 2 \left(\omega-\omega_{0}- 2 \gamma B_{0} \cos \theta \right)^{2} \right) \frac{1-\cos(qt)}{q} + \left(\omega-\omega_{0}- 2 \gamma B_{0} \cos \theta \right)^{2} t \sin (qt) \right]^{2} \right. \notag\\
        & \left. \qquad + \left(\omega-\omega_{0}- 2 \gamma B_{0} \cos \theta \right)^{2} [qt \cos (qt) - \sin(qt)]^{2}  \right\} .
    \end{align}
\end{widetext}
The Cram\'er-Rao inequality for $N$ i.i.d. quantum measurements is modified to
\begin{equation}\label{ModifiedCRB}
    \Var\left[\hat{\omega}_{0}({\bf x})\right] \geq \frac{1}{N \mathcal{F}_{\omega_{0}}^{\text{freq}}}\geq \frac{1}{N \mathcal{H}_{\omega_{0}}^{\text{freq}}}.
\end{equation}

A key task of quantum estimation theory is to saturate the lower bound of the above inequality by finding the optimal measurement scenario. Here, our measurement setup is fixed, but both the CFI and the QFI depend on the adjustable parameters of the magnetic field. Therefore, we aim to minimize the quantity 
\begin{align}\label{FreqDiff}
    \mathcal{C}_{\omega_{0}}^{\text{freq}} &=  \mathcal{H}_{\omega_{0}}^{\text{freq}} - \mathcal{F}_{\omega_{0}}^{\text{freq}} \notag\\
    &=  \left(4- \frac{1}{\rho_{00} \rho_{11}} \right) \left[\frac{\partial \rho_{00}}{\partial \omega_{0}} \right]^{2} + 4\left \vert \frac{\partial \rho_{01}}{\partial \omega_{0}} \right \vert^{2}. 
\end{align}
The inequality in \eqref{ModifiedCRB} is saturated, when the POVM is given by the projectors over the eigenvectors of $L_{\omega_{0}}$. However, in our scenario, the measurement operators are fixed, and only $L_{\omega_{0}}$ is changing, i.e., the projectors of its spectral decomposition depend on the model's parameters. Therefore, we can only have equality, when these projectors are exactly $\ket{0}\bra{0}$ and  $\ket{1}\bra{1}$. In general, this is not the case and the CFI is smaller than the QFI.

{\it Bayesian approach.} Now, $\omega_0$ is a {\it random variable} and the procedure is augmented with a {\it prior} probability distribution function (PDF) $p(\omega_0)$ with support $\Omega \subseteq \mathbb{R}$, which describes our belief before any data are obtained. After observing the data ${\bf x}$, we determine the {\it posterior} distribution function by using Bayes' theorem
\begin{equation}
\label{eq:aposteriori}
   p( \omega_0 \vert{\bf x}) =\frac{p({\bf x}\vert \omega_0) p(\omega_0)}{\int_{\Omega} p({\bf x}\vert \omega_0) p(\omega_0) \, d \omega_0},
\end{equation}
where
\begin{equation}
    p({\bf x}\vert \omega_0)=\binom{N}{k}\prod^N_{i=1} p^{x_i}(1\vert \omega_0)\left[1-p(1\vert \omega_0)\right]^{1-x_i}.
    \label{eq:PMFBayes}
\end{equation}
The posterior PDF $p( \omega_0 \vert{\bf x})$ expresses what we know about $\omega_0$ after the observation of ${\bf x}$ was done. The van Trees inequality \cite{vanTrees, Gill}
\begin{eqnarray}
E\left[ \big(\hat{\omega}_0({\bf x}) - \omega_0\big)^2 \right] \geq \frac{1}{N\int_{\Omega} \mathcal{F}_{\omega_{0}}^{\text{freq}} \, p( \omega_{0}) \,d\omega_{0} + I_{\omega_{0}}}, \label{eq:vanTrees}    
\end{eqnarray}
holds for any Bayesian estimator $\hat{\omega}_{0}({\bf x})$,
where the expectation $E[\cdot]$ is taken over the joint distribution of $\mathbf{x}$ and $\omega_0$ and
\begin{equation}\label{PriorFisher}
    I_{\omega_{0}} = \int_{\Omega} \left(\frac{\partial \ln p(\omega_{0}) }{\partial \omega_{0}}\right)^2 p(\omega_0) \, d \omega_0
\end{equation}
is the CFI of the prior PDF $p(\omega_{0})$ and thus its value depends on the choice of the prior. The derivation requires similar regularity conditions, which we discuss in Appendix \ref{AppendixB}, for both the prior PDF and the PMF in Eq. \eqref{eq:PMFBayes}. In addition, $p(\omega_0)$ must vanish at the boundaries of the set $\Omega$. The denominator on the right-hand side of the inequality in Eq. \eqref{eq:vanTrees} is upper bounded by the analog expression as in the frequentist approach, namely, the average of the QFI $\mathcal{H}_{\omega_{0}}^{\text{freq}}$ \cite{Teklu_2009}. Hence,
\begin{align}
    E\left[ \big(\hat{\omega}_0({\bf x}) - \omega_0\big)^2 \right] &\geq \frac{1}{N\int_{\Omega} \mathcal{F}_{\omega_{0}}^{\text{freq}} \, p( \omega_{0}) \,d\omega_{0} + I_{\omega_{0}}} \\ &\geq \frac{1}{N\int_{\Omega} \mathcal{H}_{\omega_{0}}^{\text{freq}}\, p(\omega_{0}) \,d\omega_{0} + I_{\omega_{0}}}, 
\end{align}
This motivates one to redefine the Bayesian version of the CFI and the QFI as
\begin{equation}\label{BayesCFI}
    \mathcal{F}_{\omega_{0}}^{\text{Bayes}} = \int_{\Omega} \mathcal{F}_{\omega_{0}}^{\text{freq}}\, p(\omega_{0}) \,d\omega_{0} + \frac{I_{\omega_{0}}}{N}
\end{equation}
and
\begin{equation}\label{BayesQFI}
    \mathcal{H}_{\omega_{0}}^{\text{Bayes}} = \int_{\Omega} \mathcal{H}_{\omega_{0}}^{\text{freq}} \,  p( \omega_{0}) \,d\omega_{0} + \frac{I_{\omega_{0}}}{N}, 
\end{equation}
respectively. Due to the cancellation of $I_{\omega_{0}}$, the Bayesian version of $\mathcal{C}_{\omega_{0}}^{\text{freq}}$ is simply
\begin{equation}\label{BayesDifference}
    \mathcal{C}_{\omega_{0}}^{\text{Bayes}} = \int_{\Omega} \mathcal{C}_{\omega_{0}}^{\text{freq}} p( \omega_{0}) \,d\omega_{0}.
\end{equation}

\subsection{Classical Data Processing}
Finally, the classical data processing techniques employed in our investigation are reviewed. We begin with the frequentist approach, and then proceed to the Bayesian approach, in which the effects of the prior are also investigated.\\

\subsubsection{Frequentist Estimators}

{\it MVU estimator.} In our scheme, binary questions are asked, and the probability of detecting a photon at time $t$ is 
\begin{eqnarray}\label{ExcitedPopulation}
    p(1;\omega_{0}) =  \vert c_{0}(t) \vert^{2}
    = \frac{4\gamma^{2}B_{0}^{2} \sin^{2} \theta}{q^{2}} \sin^{2} \left( \frac{qt}{2} \right). 
\end{eqnarray}
We observe that, upon the introduction of the notation $P_1=p(1;\omega_{0})$,
the joint PMF of the data or the likelihood function in Eq. \eqref{eq:datapmf} can be rewritten 
\begin{eqnarray}\label{FrequentistLikelihood}
    p(\mathbf{x};P_{1}) =  \binom{N}{k}P^{k}_{1} (1-P_{1})^{N-k}, 
\end{eqnarray}
where the total photon count $k=\sum_{i=1}^{N} x_{i}$ occurs during $N$ trials. It is immediate that an efficient estimator of $P_1$ exists \cite{Kay}. This is the minimum variance unbiased (MVU) estimator 
$\hat{P}_{1,\text{\tiny{MVU}}}({\bf x})$, which obeys the equation
\begin{equation}
\label{eq:MVUeq}
    \frac{\partial \ln p(\mathbf{x}; P_{1}) }{\partial P_{1}} = \frac{N}{P_1(1-P_1)} \left[\hat{P}_{1,\text{\tiny{MVU}}}({\bf x})-P_{1}\right]
\end{equation}
and is equal to $\sum_{i=1}^{N} x_{i}/N=k/N=\bar{x}$, see also Appendix \ref{AppendixB}. The quantity $\bar{x}$ is the average photon count rate obtained in the experiment. However, our task is to determine $\hat{\omega}_0({\bf x})$ and $P_1=p(1;\omega_{0})$ is a nontrivial function of $\omega_0$. The task to invert the function $p(1;\omega_{0})$ in Eq. \eqref{ExcitedPopulation} for an analytical expression of the transition frequency $\omega_{0}$ is not doable, because $\operatorname{sinc}(x)=\sin (x)/x$ is not bijective for $x \in [0,\infty)$. Still, one can solve numerically 
\begin{eqnarray}\label{eq:numerical}
    \bar{x} &=& \frac{4\gamma^{2}B_{0}^{2} \sin^{2} \theta}{q^{2}\left[\hat{\omega}_0({\bf x})\right]} \sin^{2} \left[ \frac{q\left[\hat{\omega}_0({\bf x})\right]t}{2} \right], \\
    q(\omega_0)&=&\sqrt{(\omega_{0}-\omega)^{2} + 4 \gamma B_{0} \cos \theta (\omega_{0}-\omega) + 4 \gamma^{2} B_{0}^{2}} \nonumber
\end{eqnarray}
to calculate the estimator $\hat{\omega}_{0}({\bf x})$ from the average photon count rate $\bar{x}$, but unambiguous determination of $\omega_{0}$ is only possibly under certain conditions. This will be investigated in the subsequent section containing our results. In fact, the MVU estimator approach does not guarantee in general that $\hat{\omega}_{0}({\bf x})$ inherits the properties of $\hat{P}_{1,\text{\tiny{MVU}}}({\bf x})$, namely it attains the lower bound of its Cram\'er-Rao inequality for finite $N$ and is unbiased.

{\it ML estimator.} In the maximum likelihood (ML) approach, the PMF of the data given in Eq. \eqref{FrequentistLikelihood} is maximized with respect to the unknown parameter $P_{1}$ by solving the equation
\begin{equation}\label{MLE,P}
    0=\left. \frac{\partial \ln p(\mathbf{x};P_{1})}{\partial P_{1}} \right|_{P_1=\hat {P}_{1,\text{\tiny{ML}}}({\bf x})},
\end{equation} 
which yields $\hat {P}_{1,\text{\tiny{ML}}}({\bf x})=\bar{x}$. This also means that $\hat {P}_{1,\text{\tiny{ML}}}({\bf x})=\hat {P}_{1,\text{\tiny{MVU}}}({\bf x})$. However, this time the ML approach for the unknown parameter $\omega_0$ with the PMF of the data given in Eq. \eqref{eq:datapmf} results in 
\begin{eqnarray}\label{MLE,omega}
    0&=&\left. \frac{\partial \ln p({\bf x}; \omega_0)}{\partial \omega_0} \right|_{\omega_0=\hat {\omega}_{0,\text{\tiny{ML}}}({\bf x})} \\
    &=&\left. \frac{\partial p(1;\omega_{0})}{\partial \omega_0} \frac{k-Np(1;\omega_{0})}{p(1;\omega_{0})\left[1-p(1;\omega_{0})\right]} \right|_{\omega_0=\hat {\omega}_{0,\text{\tiny{ML}}}({\bf x})}, \nonumber
\end{eqnarray}
where $p(1;\omega_{0}) \neq 0$ or $1$. The only solution is $p\big(1;\hat {\omega}_{0,\text{\tiny{ML}}}({\bf x})\big)=\bar{x}$, which is Eq. \eqref{eq:numerical}, because the maximization is performed on all allowable values of $\omega_0$
and
\begin{eqnarray}
\label{eq:2ndtest}
 &&\left. \frac{\partial^2 \ln p({\bf x}; \omega_0)}{\partial \omega^2_0} \right|_{\omega_0=\hat {\omega}_{0,\text{\tiny{ML}}}({\bf x})} \\
 &&=\frac{-N}{p(1;\omega_{0})\left[1-p(1;\omega_{0})\right]} \left[\frac{\partial p(1;\omega_{0})}{\partial \omega_0}\right]^2<0.  \nonumber
\end{eqnarray}
Contrary to the arguments brought up at the discussion of MVU estimators, here, $\hat {\omega}_{0,\text{\tiny{ML}}}({\bf x})$ is asymptotically unbiased and efficient, i.e., for $N \to \infty$ the Cram\'er-Rao bound is attained \cite{vanTrees}.\\

\subsubsection{Bayesian Estimators}

In the case of the Bayesian approach,  we incorporate our prior knowledge of the random parameter with the use of the prior probability and the Bayes' rule. The Bayesian risk is defined as $\mathcal{R}=E\left[ C\big(\hat{\omega}_0({\bf x}) - \omega_0\big) \right]$ with a cost function $C(z)$, where the expectation is taken over the joint distribution of $\mathbf{x}$ and $\omega_0$. 

{\it MMSE estimator.} When $C(z)=z^2$, the risk function $\mathcal{R}$ is minimized by the minimum mean-square error (MMSE) estimator (see Chapter $10$ in \cite{Kay}), which reads
\begin{equation}
  \hat{\omega}_{0,\text{\tiny{MMSE}}}({\bf x}) = \int_\Omega \omega_0\, p(\omega_0 \vert {\bf x}) \, d\omega_0,
\end{equation}
where the posterior PDF $p(\omega_0 \vert {\bf x})$ is obtained in Eq. \eqref{eq:aposteriori}. As the cost function is quadratic, then $\mathcal{R}$ is the Bayesian mean-square error. Thus, evaluating the Bayesian mean-square error for $\hat{\omega}_{0,\text{\tiny{MMSE}}}({\bf x})$ we obtain the minimum of $\mathcal{R}$, which is usually larger than the lower bound in the van Trees inequality, unless the posteriori PDF is Gaussian for all ${\bf x}$ \cite{vanTrees}.

{\it MAP estimator.} We choose now $C(z)=-\delta(z)$, i.e., a Dirac $\delta$ function, or the conceptually similar "hit-or-miss" cost function \cite{Kay}, whose minimization yields the maximum {\it a posteriori} (MAP) estimator:
\begin{equation}\label{eq:MAP}
    \hat{\omega}_{0,\text{\tiny{MAP}}}({\bf x}) = \arg \max_{\omega_0} \left[\ln p(\mathbf{x} \vert \omega_{0}) + \ln p(\omega_{0})\right],
\end{equation}
which is equivalent to the maximization of the logarithmic posterior PDF $\ln p( \omega_{0}\vert \mathbf{x})$. This can be seen as the Bayesian version of the ML estimator in the frequentist statistics. 

The choice of the prior PDF is crucial in Bayesian estimation, because our reasoning depends on it, and it can be done in many different ways \cite{BernardoBook, Jaynes}. Our basic stance on the matter is simple: we do not assume any knowledge with regard to the transition frequency before measurement data are taken and therefore all alternatives are equally possible. The transition frequency cannot be zero by default, because then we do not have a TLS. On the other hand the support of $p(\omega_0)$ cannot be the whole $(0,\infty)$ interval, because the single-photon detectors in our setup are unable to detect, for example, electromagnetic radiation in the frequency range of $\gamma$ rays or radio waves. Therefore, the value of $\omega_0$ lies within $\Omega=[\omega_{\text{\tiny{lower}}}, \omega_{\text{\tiny{upper}}}]$. This shows that one cannot be completely ignorant. The prior PDF used here is then the uniform distribution 
\begin{equation}\label{UniformPrior}
    p_{u}(\omega_{0}) = 
        \begin{cases}
        \frac{1}{\omega_{\text{\tiny{upper}}}-\omega_{\text{\tiny{lower}}}}, \quad &\text{if}\,\,\, \omega_0 \in \Omega, \\
        0,  &\text{otherwise.}
        \end{cases}
\end{equation}

In case the parameter space $\Omega$ is a continuum (say, an interval), the uniform prior suffers from the lack of invariance from differentiable transformation. In other words, noninformative priors should presumably be invariant under reparametrization. The Jeffrey's prior addresses this issue and is constructed based on the frequentist CFI $\mathcal{F}_{\omega_{0}}^{\text{freq}}$ as \cite{Jeffrey, Berger} 
\begin{eqnarray}
    &&p_{J}(\omega_{0}) = \sqrt{\mathcal{F}_{\omega_{0}}^{\text{freq}}}/\int_{\Omega} \sqrt{\mathcal{F}_{\omega_{0}}^{\text{freq}}} d\omega_{0} \\
    &&\quad \text{with} \quad \int_{\Omega} \sqrt{\mathcal{F}_{\omega_{0}}^{\text{freq}}} d\omega_{0}<\infty \nonumber 
\end{eqnarray}
It is worth noting that the Jeffrey's prior can be improper, such that it cannot be normalized to one. However, we do not have this problem here. Now, making use of Eq. \eqref{CFI} and the relation $p(1\vert\omega_{0})=|c_0(t)|^2$, we find that the Jeffrey's prior is proportional to the Beta function, i.e.,
\begin{align}\label{posterior}
    p_{J}(\omega_0) &\propto  \Big\{p(1\vert\omega_{0}) \left[1-p(1\vert\omega_{0})\right]\Big\}^{-\frac{1}{2}} \left|\frac{\partial p(1\vert\omega_{0})}{\partial \omega_0} \right|.
\end{align}
This is a proper prior and does not give rise to improper posterior.

To demonstrate the effect of an informative prior, i.e., some knowledge on the transition frequency is given, we also consider a Gaussian prior
\begin{align}
    p_{G}(\omega_{0})= \frac{1}{\sqrt{2\pi} \sigma} e^{-\frac{(\omega_{0} - \omega_{\text{\tiny{mean}}})^2}{2\sigma^{2}}},
\end{align}
where $\omega_{\text{\tiny{mean}}}$ denotes the value of $\omega_{0}$ we expect before the measurements and $\sigma$ describes our uncertainty about our expectations. 

In the following section, we investigate all the introduced estimators in both the frequentist and Bayesian approaches. Furthermore, properties of QFI and CFI including their difference are discussed.\\

\section{Results}
\label{4}

In this section, we conduct numerical investigations on the quantities discussed in Sec. \ref{3}. In the first part, we investigate the effects of the control parameters of the gyrating magnetic field on the frequentist CFI and its deviation from its upper bound, the frequentist QFI. Then, we proceed to analyze the ML estimator. In the second part, we turn to Bayesian statistics. We first investigate the Bayesian version of CFI and its deviation from the QFI. The MMSE and MAP estimators are studied using various informative and noninformative priors. To make the equations dimensionless, we absorb the time under which the system evolves into the definition of dimensionless quantities, and by doing it systematically it leads to an implicit time dependence of the model.

\begin{figure}[t!]
\includegraphics[width=0.5\textwidth]{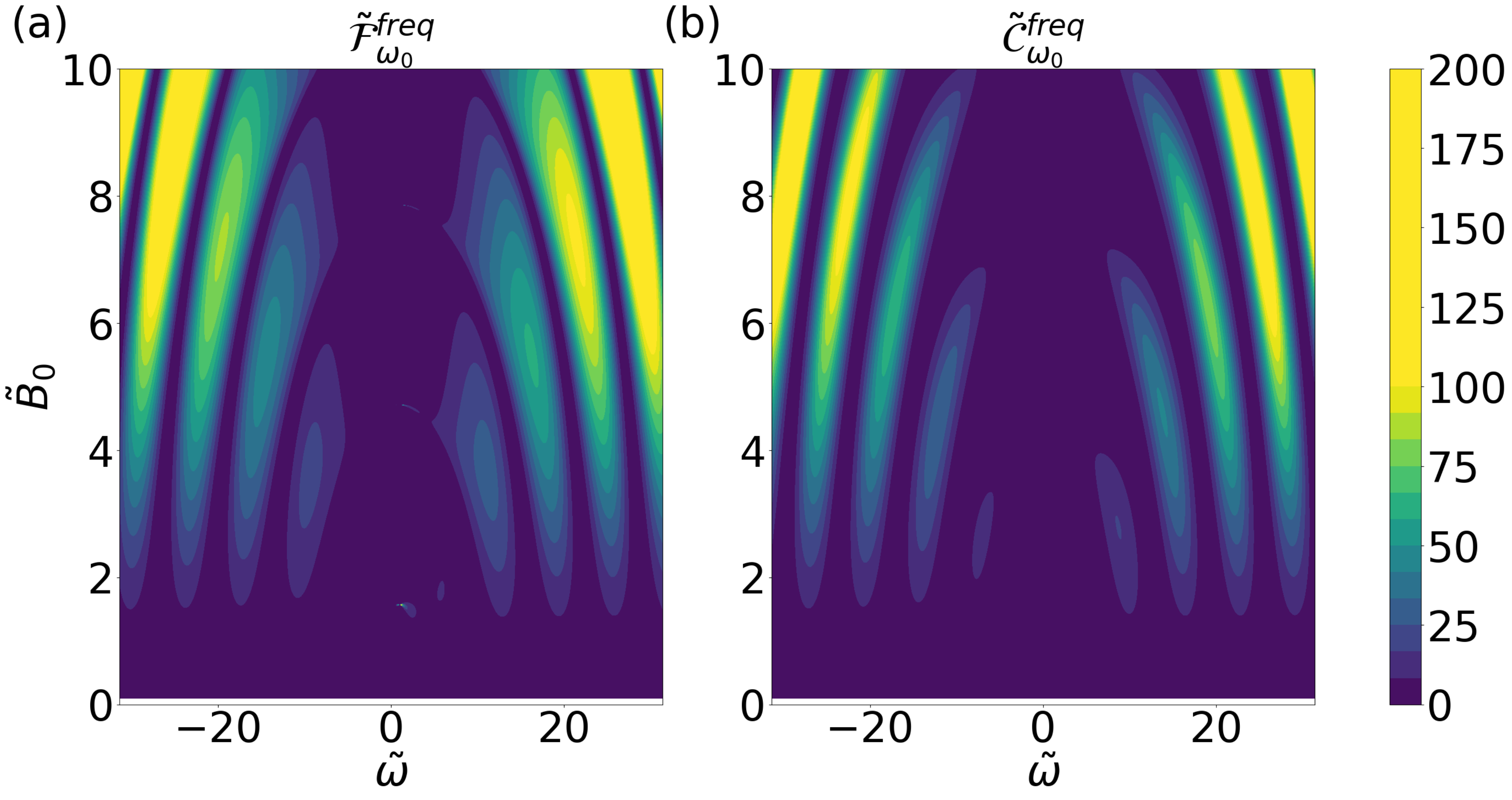}
\includegraphics[width=0.48\textwidth]{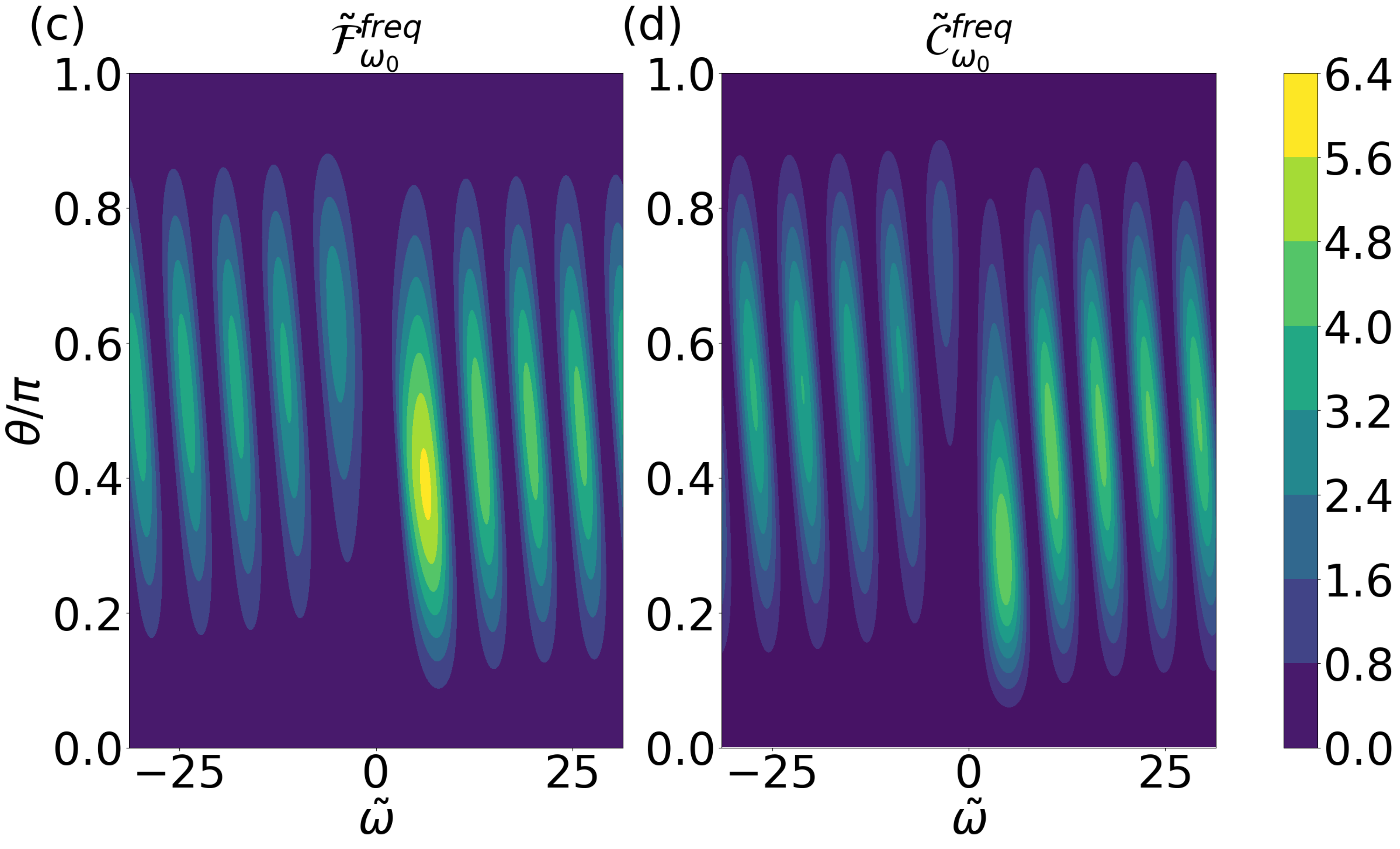}
\includegraphics[width=0.48\textwidth]{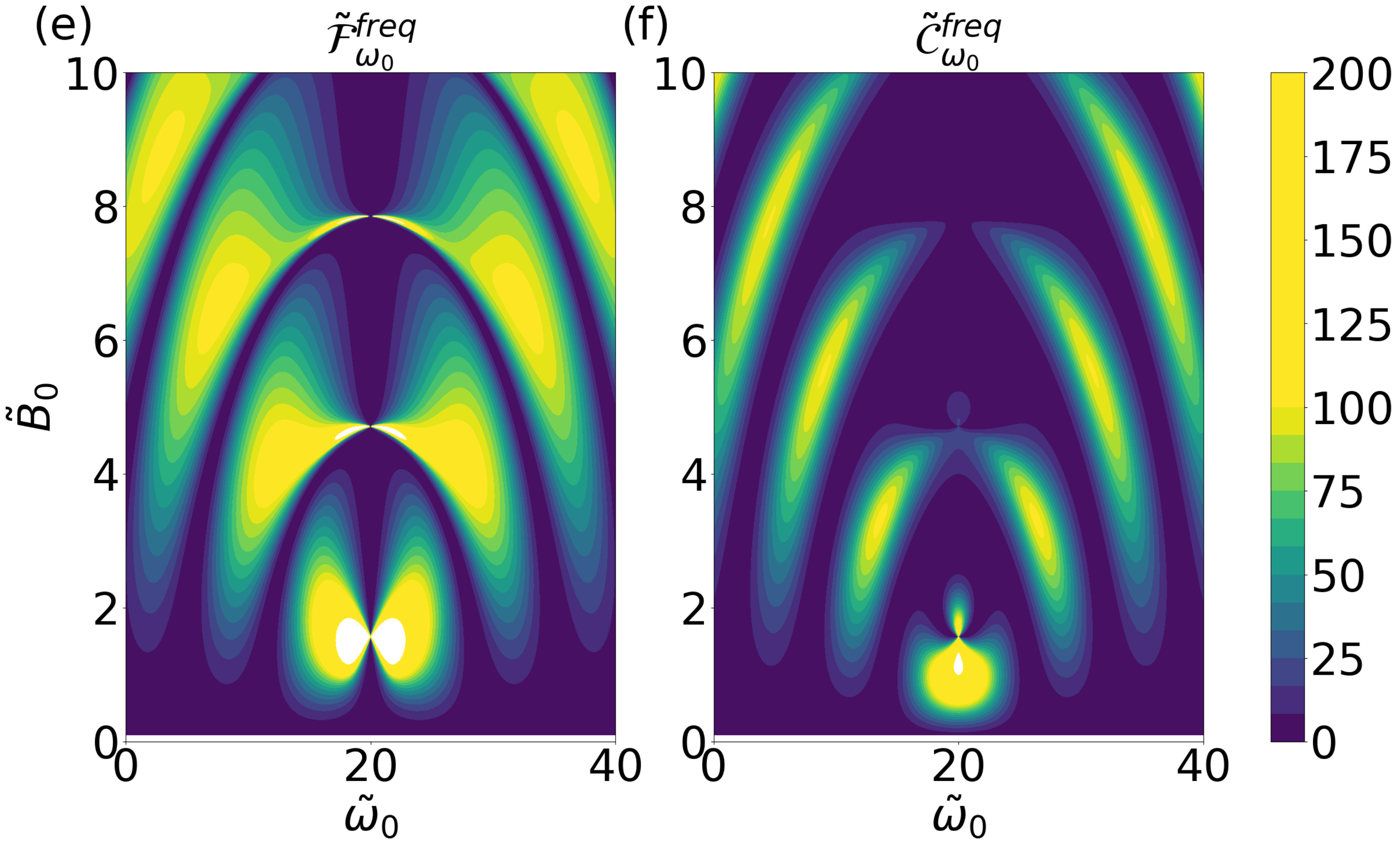}


    \caption{Contour plots of CFI $\tilde{\mathcal{F}}_{\omega_{0}}^{\text{freq}}$ and its deviation $\tilde{\mathcal{C}}_{\omega_{0}}^{\text{freq}}$ from QFI. As a demonstration, but keeping in mind that $\tilde{\omega}_{0}$ is unknown, we make the assumption $\tilde{\omega}_{0}=1$. (a) $\tilde{\mathcal{F}}_{\omega_{0}}^{\text{freq}}$ and (b) $\tilde{\mathcal{C}}_{\omega_{0}}^{\text{freq}}$ upon the variation of the magnetic-field strength $\tilde{B}_{0}$ and its angular speed $\tilde{\omega}$. We set $\theta = \pi/2$. In regions such as $[-30,-20] \times [6.8, 10]$ and $[20,30] \times [6.5, 10]$, large values of $\tilde{\mathcal{F}}_{\omega_{0}}^{\text{freq}}$ are attained while $\tilde{\mathcal{C}}_{\omega_{0}}^{\text{freq}}$ remain small. (c) $\tilde{\mathcal{F}}_{\omega_{0}}^{\text{freq}}$ and (d) $\tilde{\mathcal{C}}_{\omega_{0}}^{\text{freq}}$ upon the variation of the orientation of the magnetic field $\theta$ and its angular speed $\tilde{\omega}$, where $\tilde{B}_{0} = 1$. (e) $\tilde{\mathcal{F}}_{\omega_{0}}^{\text{freq}}$ and (f) $\tilde{\mathcal{C}}_{\omega_{0}}^{\text{freq}}$ upon the variation of $\tilde{B}_{0}$ and $\tilde{\omega}_{0}$, where $\tilde{\omega} = 20$ and $\theta = \pi/2$. In general, there are different areas in the figures, where either CFI is small or its deviation from QFI is large, which are not optimal for the estimation of $\omega_0$.} 

    \label{fig:FI}
\end{figure}

\subsection{Frequentist Statistical Analysis}
{\it Frequentist Fisher Information.} In the frequentist approach, research articles \cite{Genoni, Pezze, Razzoli, Sanavio} often focus also on comparative measures between the CFI and the QFI. However, it is rarely stated, e.g., see  Ref. \cite{Tsang} ( Ref. [$9$] of the paper), that the minimization of these measures may not always be ideal for the accuracy of the estimated parameters. While the CFI can be brought close to its upper bound, it can still be very small, which is a bad scenario for achieving high accuracy in the classical postprocessing of the data. Therefore, our first task is to examine the CFI together with its deviation from the QFI for the TLS described in Sec. \ref{2}. We have three control parameters, namely, the magnetic-field strength $B_{0}$, the angle $\theta$, and the angular speed $\omega$. We employ dimensionless quantities in our investigation. To this end, we define the dimensionless quantities $\tilde{\omega}_{0} = \omega_{0} t$, $\tilde{\omega} = \omega t$, $\tilde{B}_{0} = \gamma B_{0} t$, $\tilde{q} = qt$, $\tilde{\omega}_{\text{\tiny{mean}}}=\omega_{\text{\tiny{mean}}}t$, and $\tilde{\sigma}=\sigma t$. We further define the dimensionless CFI and the dimensionless QFI as $\tilde{\mathcal{F}}_{\omega_{0}}^{\text{freq}} = \omega^{2}\mathcal{F}_{\omega_{0}}^{\text{freq}}$ and $\tilde{\mathcal{H}}_{\omega_{0}}^{\text{freq}} = \omega^{2}\mathcal{H}_{\omega_{0}}^{\text{freq}}$. Since there are three control parameters, we study the behavior of $\tilde{\mathcal{F}}_{\omega_{0}}^{\text{freq}}$ and $\tilde{\mathcal{C}}_{\omega_{0}}^{\text{freq}}$ for different values of the dimensionless transition frequency $\tilde{\omega}_{0}$ for the scenarios in which (i) we vary $\tilde{B}_{0}$ and $\tilde{\omega}$ while holding $\theta$ fixed, (ii) we vary $\theta$ and $\tilde{\omega}$ while holding $\tilde{B}_{0}$ fixed, or (iii) we vary $\tilde{B}_{0}$ and $\theta$ while holding $\tilde{\omega}$ fixed. The goal is to find out the optimal accuracy for different possible values of the transition frequency $\tilde{\omega}_{0}$. It is important to keep in mind that $\tilde{\omega}_{0}$ is an unknown quantity, and the time $t$ is determined by the experiment. Thus, any fixed value of $\tilde{\omega}_{0}$ represents a possible case with only a demonstrative role. The following aims to provide an intuition how $\tilde{\mathcal{F}}_{\omega_{0}}^{\text{freq}}$ and $\tilde{\mathcal{C}}_{\omega_{0}}^{\text{freq}}$ respond to the changes of the control parameters. 

Figs. \ref{fig:FI}(a) and \ref{fig:FI}(b) show the contour plots of $\tilde{\mathcal{F}}_{\omega_{0}}^{\text{freq}}$ and $\tilde{\mathcal{C}}_{\omega_{0}}^{\text{freq}}$, respectively, upon the variation of $\tilde{\omega}$ and $\tilde{B}_{0}$, while we let $\tilde{\omega}_0=1$ and $\theta$ is fixed at $\pi/2$ such that the magnetic field is constrained to rotate in the $x$-$y$ plane. We also allow $\tilde{\omega}$ to take negative values so that we can observe the effects of the rotation direction of the magnetic field on $\tilde{\mathcal{F}}_{\omega_{0}}^{\text{freq}}$ and $\tilde{\mathcal{C}}_{\omega_{0}}^{\text{freq}}$.  These figures demonstrate that in most regions of small deviations between the CFI and the QFI, the CFI has very small values. Performing estimation using magnetic field of such control parameters is detrimental, since the accuracy of the estimators is determined by the CFI. However, while the graphs of  $\tilde{\mathcal{F}}_{\omega_{0}}^{\text{freq}}$ and $\tilde{\mathcal{C}}_{\omega_{0}}^{\text{freq}}$ share significant similarities, their regions of large values are not identical. It is therefore possible to locate control parameters for which the CFI attains large values, while its deviation from the QFI is small. This takes place, for example, in the regions $(\tilde{\omega}, \tilde{B}_{0})\in [-30,-20] \times [6.8,10]$ or $[20,30] \times [6.5,10]$ in Figs. \ref{fig:FI}(a) and Fig. \ref{fig:FI}(b). In Figs. \ref{fig:FI}(c) and \ref{fig:FI}(d) the effects of varying $\theta$ and $\tilde{\omega}$ on $\tilde{\mathcal{F}}_{\omega_{0}}^{\text{freq}}$ and $\tilde{\mathcal{C}}_{\omega_{0}}^{\text{freq}}$ are shown respectively, while holding $\tilde{B}_{0}$ and $\tilde{\omega}_0$ constant. Similar to the previous case, there exist many regions in which both the CFI and the QFI attain small values. However, it is still possible again to locate the control parameters settings with which the CFI attains large values while the deviation from the QFI is small. Finally, Fig. \ref{fig:FI}(e) shows the CFI by varying $\tilde{B}_{0}$ and $\tilde{\omega}_0$, while holding the rest of the parameters constant. The accuracy is lower when
the system experiences a weaker magnetic field. Now, comparing Fig. \ref{fig:FI}(e) with Fig. \ref{fig:FI}(f), we find again regions, where $\tilde{\mathcal{F}}_{\omega_{0}}^{\text{freq}}$ is large and $\tilde{\mathcal{C}}_{\omega_{0}}^{\text{freq}}$ is small. In the other cases, we have found no new features, and therefore these results are not presented. 

Given these difficulties, the following steps may be relevant to improving the accuracy of the estimation.
\begin{enumerate}[(i)]
    \item Determining the ranges of the model’s parameters: the experimentalist has a prior guess on $\omega_0$, namely, an interval of possible values, which is based on either former experiments or theoretical models of the material; the rest of the parameters also have constraints due to the experimental instruments. Therefore, we omit further discussion of this step.

    \item Optimal control parameters: a numerical analysis of $\tilde{\mathcal{F}}_{\omega_{0}}^{\text{freq}}$ and $\tilde{\mathcal{C}}_{\omega_{0}}^{\text{freq}}$ for these constrained values are carried out; the search for $\tilde{\omega}$, $\tilde{B}_0$, and $\theta$ has to ensure large values of $\tilde{\mathcal{F}}_{\omega_{0}}^{\text{freq}}$ for the guessed values of $\omega_0$, while $\tilde{\mathcal{C}}_{\omega_{0}}^{\text{freq}}$ is kept as small as possible.
    
    \item Estimation: collect data and then evaluate the ML estimator.
\end{enumerate}

\begin{figure}[t!]
       \includegraphics[width=0.4\textwidth]{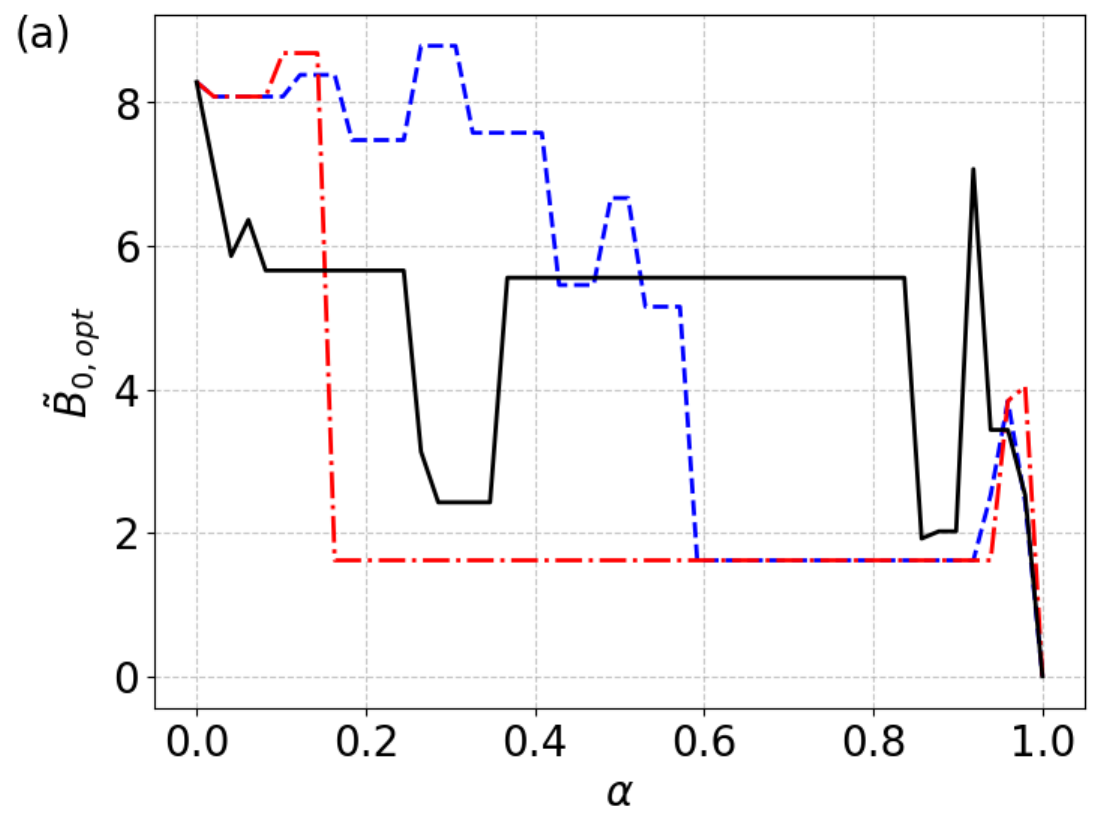}
       \includegraphics[width=0.4\textwidth]{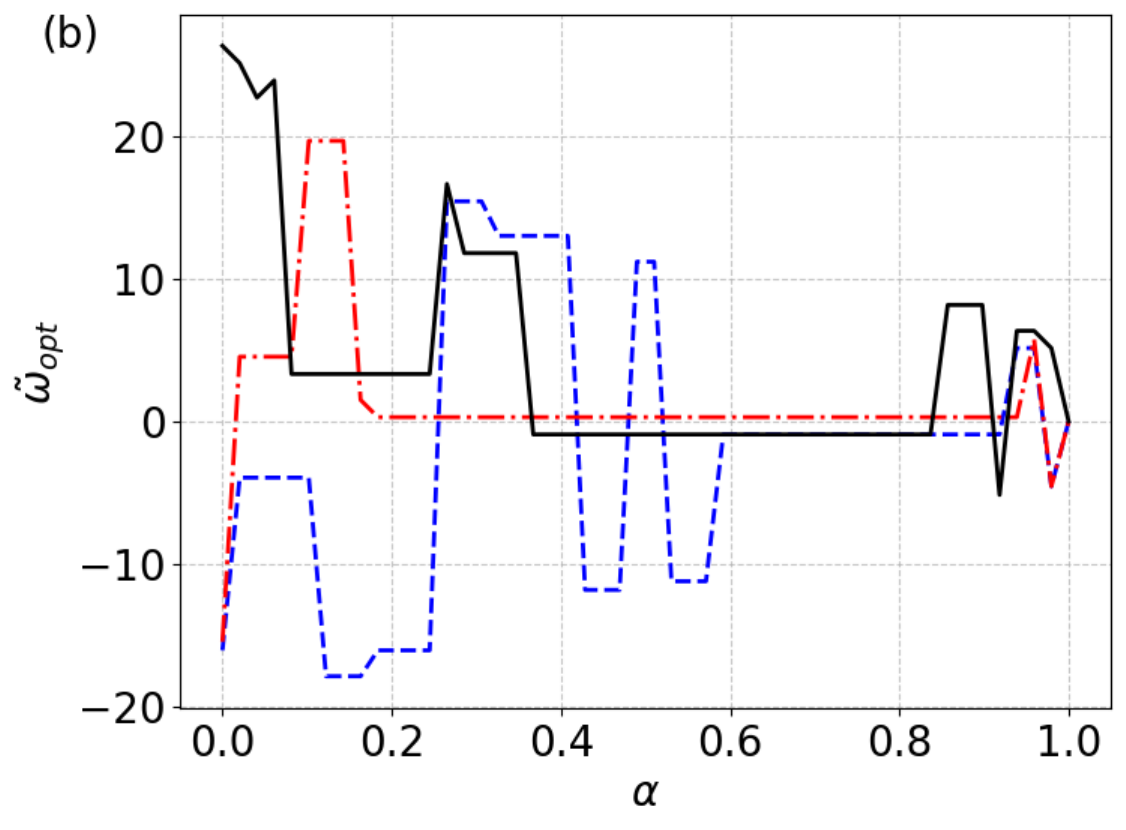} 
       \includegraphics[width=0.4\textwidth]{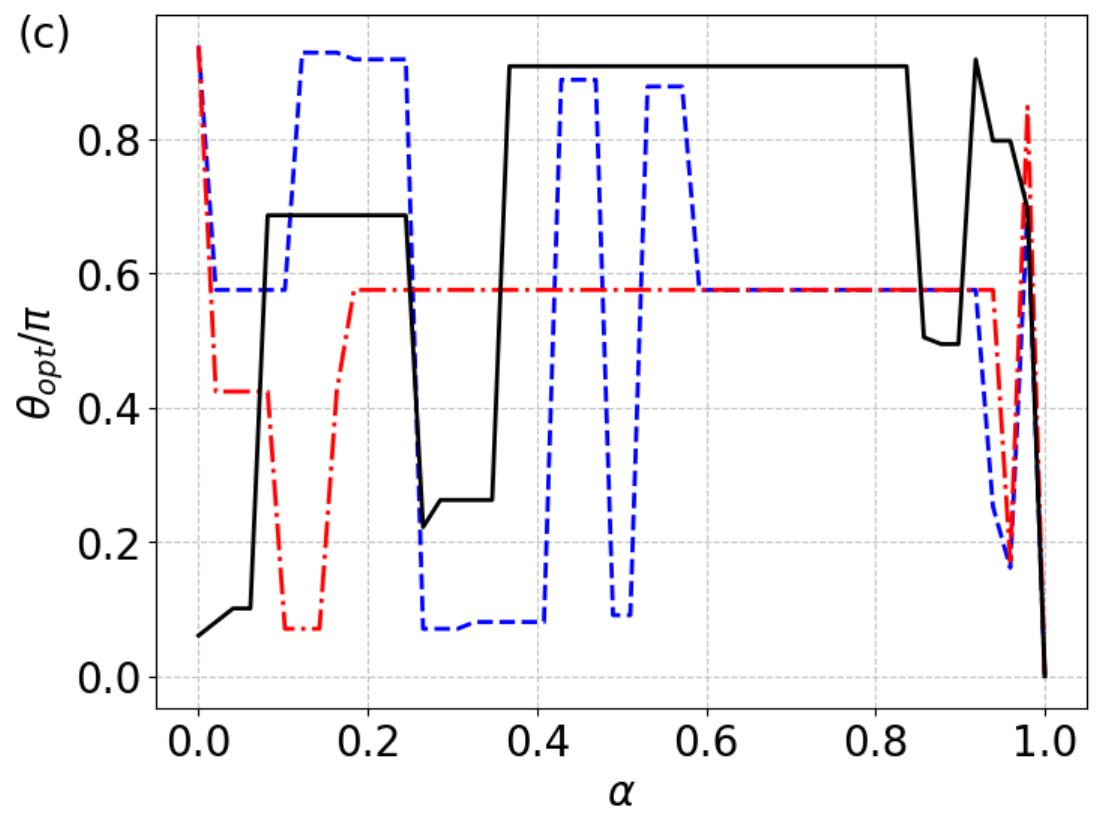}     
       
   \caption{Optimal control parameters of the magnetic field for different $\tilde{\omega}_{0}$ and $\alpha$. The search was performed on the range $\tilde{B}_{0} \in [0,10]$ ($100$ points), $\tilde{\omega} \in [-30,30] $($100$ points) and $\theta \in [\epsilon, \pi - \epsilon] $($100$ points), where $\epsilon=0.0001$ serves as a regularization factor. In the case of $\alpha$, we used $50$ points, and the resulted optimal values are connected by a line. The cases $\tilde{\omega}_{0}=0.1$, $\tilde{\omega}_{0}=1$ and $\tilde{\omega}_{0}=10$ are represented by the blue dashed line, the red dash-dotted line and the black solid line respectively. (a) Optimal magnetic-field strength $\tilde{B}_{0,\text{opt}}$, (b) optimal angular speed $\tilde{\omega}_{\text{opt}}$, and (c) optimal angle $\theta_{\text{opt}}$ as functions of $\alpha$.}

   \label{fig:Opt}
\end{figure}

{\it Optimal control parameters.} Since we have two objectives, namely, the maximization of the frequentist CFI and the minimization of its deviation from the frequentist QFI, it is useful to define the following cost function
\begin{equation}
\label{eq:costalphafreq}
    C^{\text{freq}}(\alpha) = \alpha \tilde{\mathcal{C}}_{\omega_{0}}^{\text{freq}}(\tilde{B}_{0}, \tilde{\omega}, \theta, \tilde{\omega}_{0}) +  \frac{(1 - \alpha)}{\tilde{\mathcal{F}}_{\omega_{0}}^{\text{freq}}(\tilde{B}_{0}, \tilde{\omega}, \theta, \tilde{\omega}_{0})} ,
\end{equation}
where $\alpha \in [0, 1]$ serves as the weight to quantify the relative importance of $1/\tilde{\mathcal{F}}_{\omega_{0}}^{\text{freq}}(\tilde{B}_{0}, \tilde{\omega}, \theta, \tilde{\omega}_{0})$, i.e., the accuracy of the classical data processing, and $\tilde{\mathcal{C}}_{\omega_{0}}^{\text{freq}}(\tilde{B}_{0}, \tilde{\omega}, \theta, \tilde{\omega}_{0})$ in the optimization problem. We perform a search on the control parameters of the magnetic field $\tilde{B}_{0}$, $\tilde{\omega}$ and $\theta$ such that the above cost function attains its minimum for the values of $\tilde{\omega}_{0}$ and $\alpha$ we choose. The optimal parameters of the gyrating magnetic field are denoted as $\{\tilde{B}_{0,\text{opt}}, \tilde{\omega}_{\text{opt}}, \theta_{\text{opt}}\}$. The optimized parameters of the magnetic field are shown in Fig. \ref{fig:Opt}. These results indicate that certain optimal values are unaffected by the change of $\alpha$, i.e., $C^{\text{freq}}(\alpha)$ obtains the same minimal value. In fact, we are searching for the optimal parameters discussed in Fig. \ref{fig:FI}. It is worth mentioning that the numerical search keeps only the first found optimum, i.e., the values of the control parameters corresponding to the minimum value of the cost function are not unique. Thus, other optimal solutions are not presented. This is the reason why at $\alpha=1$ all the optimal values are zero, because in that case $\tilde{\mathcal{C}}_{\omega_{0}}^{\text{freq}}(0, 0, 0, \tilde{\omega}_{0})=0$. We emphasize again that this investigation is only for demonstrative purposes due to the fact the $\tilde{\omega}_{0}$ is the unknown quantity subject to estimation, and the ranges of the control parameters are subject to experimental constraints. However, if an interval of possible values of the transition frequency is identified, then for a fixed value of $\alpha$ the optimal parameters $\{\tilde{B}_{0,\text{opt}}, \tilde{\omega}_{\text{opt}}, \theta_{\text{opt}}\}$ have to be found such that $C^{\text{freq}}(\alpha)$ is small for all possible values of $\tilde{\omega}_0$. A solution is to calculate the average value of $C^{\text{freq}}(\alpha)$ over the identified interval of $\tilde{\omega}_0$ and then search for the optimal parameters. This strategy is similar to the Bayesian approach with uniform prior PDF, which will be discussed later in the section. The readers are reminded that the search is performed on a selected range of values of the control parameters $\{\tilde{B}_{0}, \tilde{\omega}, \tilde{\theta}\}$. Due to this fact, the values of the control parameters corresponding to the global minimum of the cost function may lie beyond these regions, although they might not be reachable from an experimental point of view.

{\it Maximum Likelihood Estimator.} We now demonstrate, how the ML estimator of $\tilde{\omega}_{0}$ is evaluated. The subsequent discussion is independent of the optimal values of the control parameters and only offers insight into the qualitative features of the estimation. The evaluation of the ML estimator requires the inversion of \eqref{eq:numerical}. The Newton's method is employed to find the preimage of the unnormalized sinc function $\sinc x=\sin x/x$, where $x \in \mathbb{R}$. The fact that the function $\sinc x$ is not injective implies that its inverse does not exist in general. However, we may take advantage of the domain between the central local maximum at $x=0$ and the first root of $\sinc x$ at $\pi$. In this interval $[0,\pi]$, the inversion of \eqref{eq:numerical} is possible, when $\sinc x$ takes values between zero and one, which leads to the constraint 
\begin{equation}\label{Validity}
 1 \geq \frac{\sqrt{\bar{x}}}{\tilde{B}_{0}|\sin \theta|} \geq 0.  
\end{equation}
It is worth nothing that the right-hand side inequality is automatically fulfilled. We need to assert another condition 
\begin{equation}
  S^{2}- \tilde B_{0}^{2}  \sin^{2} \theta > 0 
  \label{eq:complexroot}
\end{equation}
with
\begin{equation}
 S = \sinc^{-1} \left( \frac{\sqrt{\bar{x}}}{\tilde B_{0} |\sin \theta|} \right), 
\end{equation}
such that the estimates of $\tilde{\omega}_0$ are real and distinct.

\begin{figure}[t!]
        \includegraphics[width=0.5\textwidth]{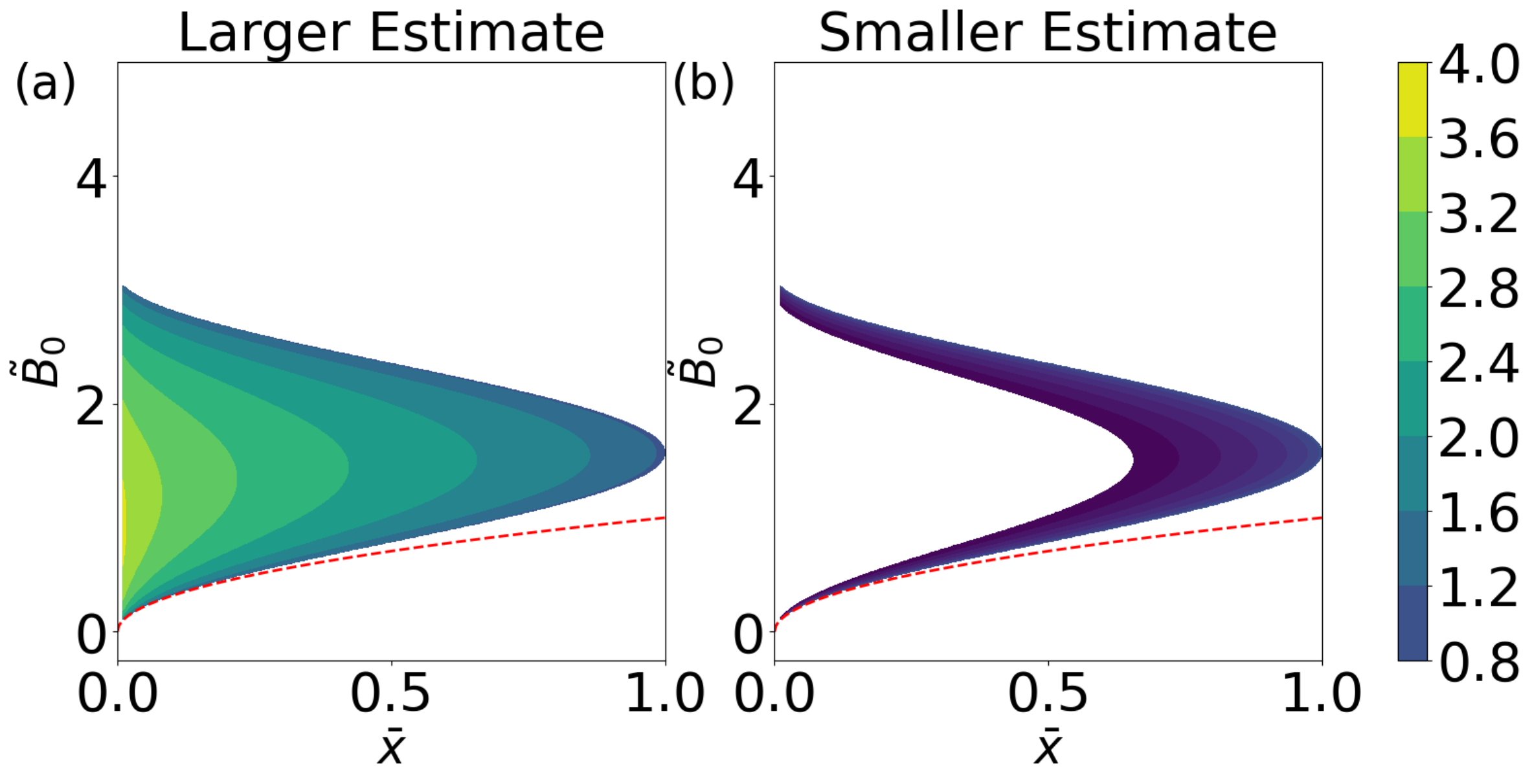}
        \includegraphics[width=0.48\textwidth]{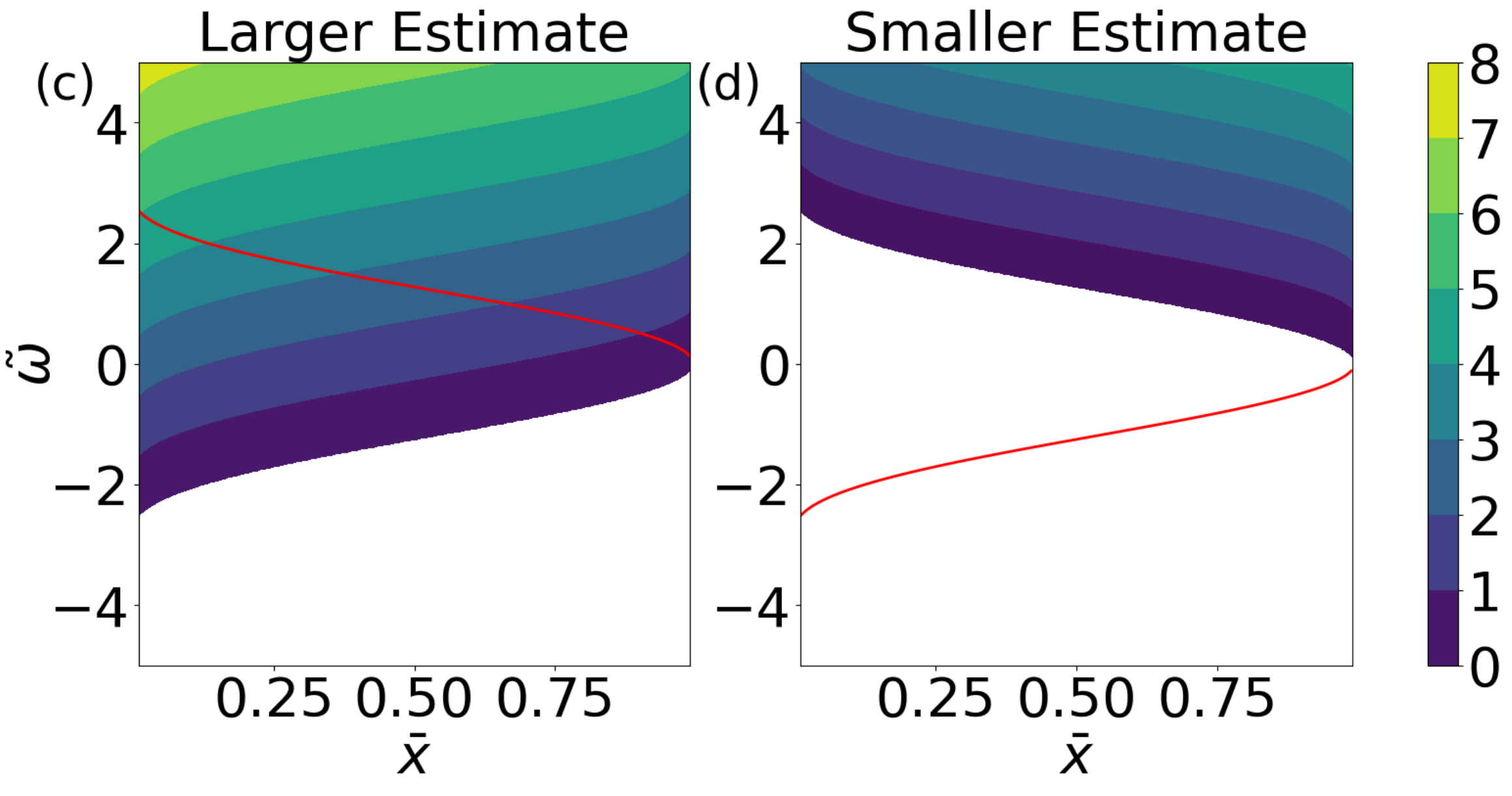}
    \caption{Maximum likelihood (ML) estimator: (a) Larger estimates and (b) smaller estimates of $\tilde{\omega}_{0}$ as functions of the magnetic-field strength $\tilde{B}_{0}$ and the average photon count rate $\bar{x}$. The magnetic field rotates at $\tilde{\omega}=1$ in the $x$-$y$ plane. The dotted red line represents the condition \eqref{Validity}, when the equality holds, i.e., $\tilde{B}_{0} = \frac{\sqrt{\bar{x}}}{\vert \sin \theta \vert}$. Above this dotted red line $\sinc^{-1} \left( \frac{\sqrt{\bar{x}}}{\tilde B_{0} |\sin \theta|} \right)$ is well-defined. White regions indicate either negative- or complex-valued estimates. (c) Larger estimates and (d) smaller estimates as functions of the angular speed $\tilde{\omega}_{0}$ and the average photon count rate $\bar{x}$. The solid red line indicates positions at which the smaller-valued estimate attains zero in the left figure or the larger-valued estimate is zero in the right figure. The estimates attain negative values in the white region.}
    \label{fig:QF_Roots}
    
\end{figure}

After ensuring that the inverse of $\sinc x$ can be evaluated, the inversion of \eqref{eq:numerical} leads to two possible solutions of $\hat{\tilde{\omega}}_{0}({\bf x})$ or estimates of $\tilde{\omega}_0$. Two possible scenarios are (i) either one estimate is positive while the other is negative, or (ii) both estimates are positive. In case (i), we reject the negative-valued estimate since $\hat{\tilde{\omega}}_{0}({\bf x})$ admits positive values only. If case (ii) occurs, the control parameter has to be adjusted until one estimate becomes negative. The two estimates of $\tilde{\omega}_{0}$ as functions of the control parameters are shown in Fig. \ref{fig:QF_Roots} and only regions in which the  estimates admit positive values are displayed. We constrained the magnetic field to the $x$-$y$ plane. Figs. \ref{fig:QF_Roots}(a) and (b) depict the two estimates with respect to $\tilde{B}_{0}$ and $\bar{x}$ upon holding $\tilde{\omega}$ constant. The first observation is that certain regions are forbidden for $\hat{\tilde{\omega}}_{0}({\bf x})$, because we have obtained complex values, i.e., the condition in \eqref{eq:complexroot} is not satisfied. An example is $\theta=\pi/2$ and $\tilde{B}_0>\pi$, which can be seen in Figs. \ref{fig:QF_Roots}(a) and (b), the colored regions at $\bar{x}=0$ start at $\tilde{B}_0=\pi$. The origin of the condition \eqref{eq:complexroot} is due to the chosen range $[0,\pi]$ of the $\sinc^{-1}(x)$. This means that for large values of the dimensionless magnetic-field amplitude $\tilde{B}_0$ either the angle $\theta$ has to be set to a value such that \eqref{eq:complexroot} is valid or a different interval has to be chosen for the inversion of  $\sinc(x)$. Furthermore, the region in which the smaller estimate admits negative values has to be discarded since $\tilde{\omega}_{0} \in \mathbb{R}^{+}$. 

One can observe that $\hat{\tilde{\omega}}_{0}({\bf x})$ may admit two possible values in certain regions. This feature arises due to the nonlinear relation between $\tilde{\omega}_{0}$ and $\bar{x}$. However, it is possible to tune the control variables so that the smaller estimate admits only negative values and hence it can be rejected. In particular, as $\tilde{\omega}$ is decreased, the smaller estimate decreases in value. Once the threshold for $\tilde{\omega}$ is reached, the smaller estimate is zero for all values of $\bar{x}$ and $\tilde{B}_{0}$ and hence can be rejected. We also investigated the effects of different angular speed $\tilde{\omega}$ of the magnetic field and average photon count rate $\bar{x}$ at constant $\tilde{B}_{0}$. The result is shown in Figs. \ref{fig:QF_Roots}(c) and (d). The unambiguous estimate scenario occurs in the colored region below the red line, in Fig. \ref{fig:QF_Roots}(c). However, there is price to get unambiguous estimates, because there is an upper limit on the range of the transition frequency.

As mentioned in Sec. \ref{3}, the efficiency of the estimator for finite data points is lost due to the nonlinear relation \eqref{eq:numerical} between $\omega_{0}$ and $\bar{x}$. Therefore, it is required to collect large number of data points so that the asymptotic property ensures that the dimensionless estimator $\hat{\tilde{\omega}}_{0}({\bf x})$ becomes efficient. However, collecting a sufficiently large set of data required for the asymptotic property within a given time frame may be subject to experimental constraints.  The required number of observations for a given level of accuracy can be reduced by using the first estimate of $\tilde{\omega}_{0}$ to search for a new set of optimal control parameters which minimize the cost function. If these set of parameters lead to a larger value of $\tilde{\mathcal{F}}_{\omega_{0}}^{\text{freq}}$, then a new experiment is set up and carried out by taking into account the unambiguity issues of the ML estimation, but this is doable because multiple optimal parameters are available. 

\begin{figure}[t!]
    \includegraphics[width=0.58\textwidth]{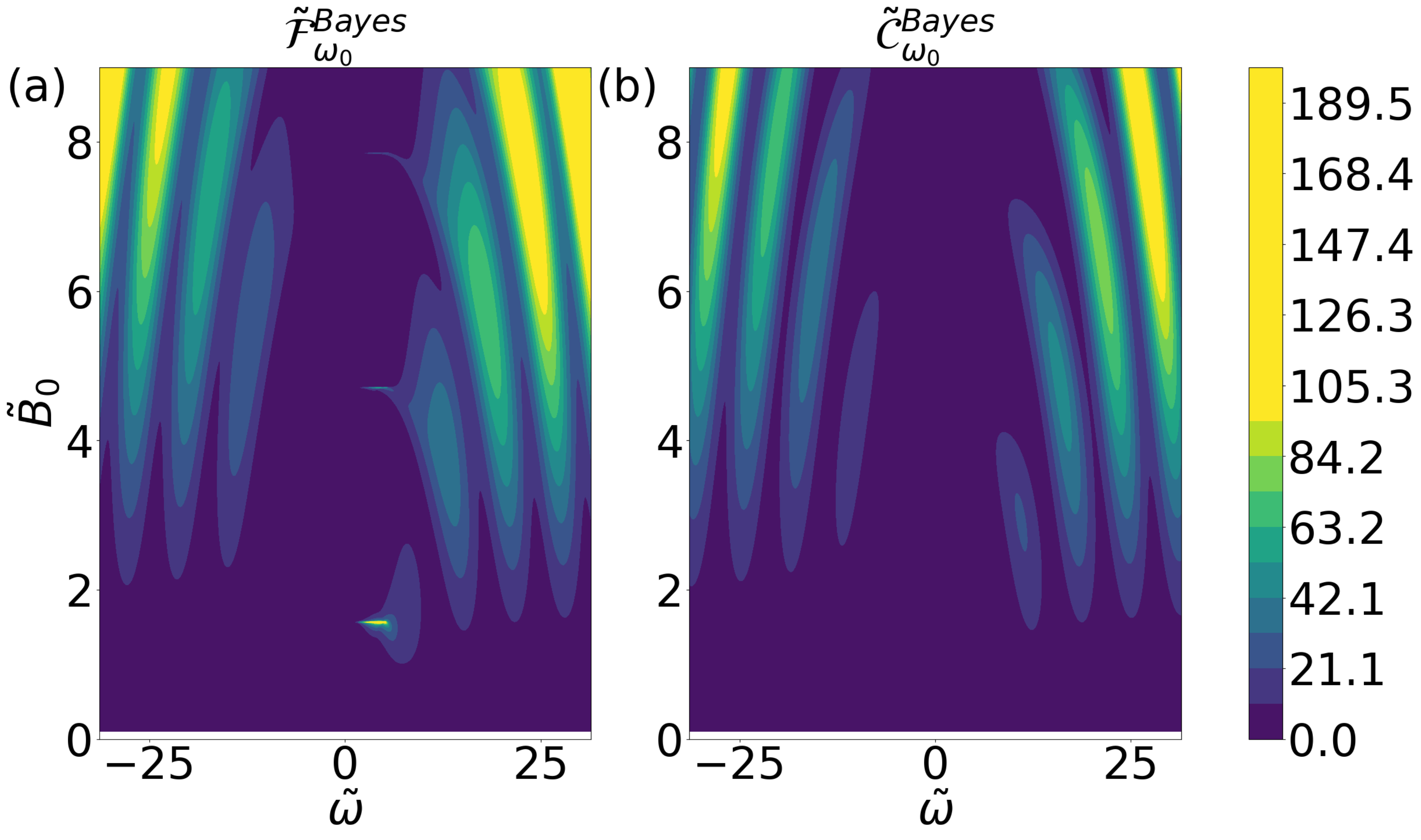}

    \caption{Contour plots of $\tilde{\mathcal{F}}_{\omega_{0}}^{\text{Bayes}}$ and $\tilde{\mathcal{C}}_{\omega_{0}}^{\text{Bayes}}$. (a) $\tilde{\mathcal{F}}_{\omega_{0}}^{\text{Bayes}}$ and (b) $\tilde{\mathcal{C}}_{\omega_{0}}^{\text{Bayes}}$ for the scenario in which the magnetic field is rotating in the $x$-$y$ plane at different angular speed $\tilde{\omega}$. We used the uniform prior distribution $p(\omega_0)$ in Eq. \eqref{UniformPrior} with $\tilde{\omega}_{\text{\tiny{lower}}}=1.5$ and $\tilde{\omega}_{\text{\tiny{upper}}}=5$.}

    \label{fig:Bayesian}
\end{figure}

\begin{figure}[t!]
    \includegraphics[width=0.44\textwidth]{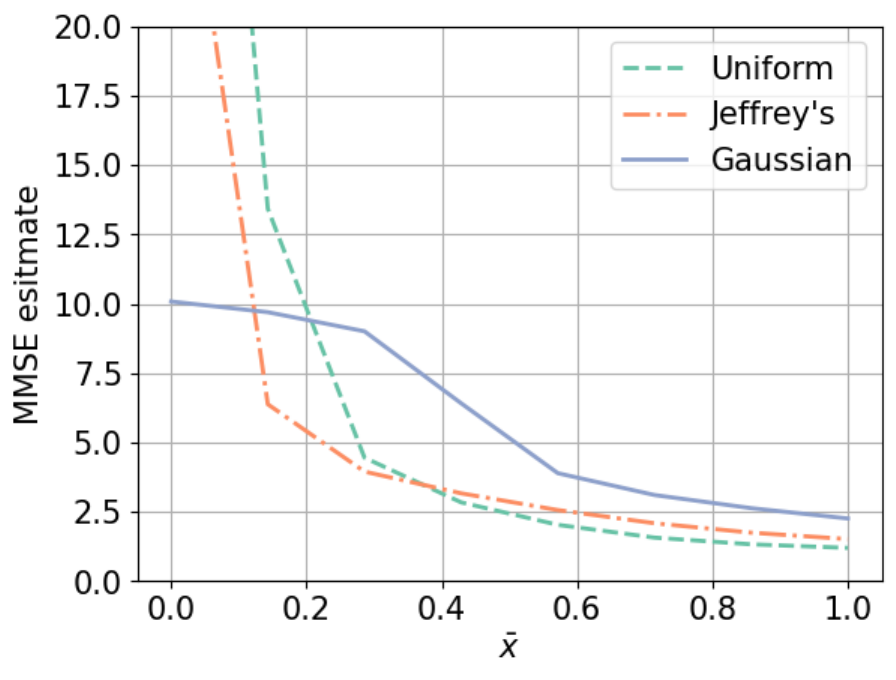}
    \caption{Minimum mean-square error (MMSE) estimator as functions of the average photon count rate $\bar{x}$ with the use of the uniform prior, the Jeffrey's prior and Gaussian prior.The parameters of the magnetic field used are $\tilde{\omega} = 1$, $\tilde{B}_{0} = 1$, and $\theta = \pi/2$, where $N=8$ data points are collected. The lower and upper limits of the numerical integration are $\tilde{\omega}_{\text{lower}} = 0.1$ and $\tilde{\omega}_{\text{upper}}=100$. The mean and the standard deviation of the Gaussian prior are chosen to be $\tilde{\omega}_{\text{\tiny{mean}}}=10$ and $\tilde{\sigma} = 2$, respectively.} 
    \label{fig:MSSE}
\end{figure}

\subsection{Bayesian statistical analysis}

{\it Bayesian Fisher Information.} We then turn to investigating the behavior of the Bayesian CFI $\mathcal{F}_{\omega_{0}}^{\text{Bayes}}$
and its deviation $\mathcal{C}_{\omega_{0}}^{\text{Bayes}}$ from the Bayesian QFI, which are defined in Eqs. \eqref{BayesCFI} and \eqref{BayesQFI}, respectively. Similar to the numerical investigations of the frequentist counterparts, we study the dimensionless version of these quantities. Due to the difficulty in the analytical calculations of the integrals in \eqref{BayesCFI} and \eqref{BayesQFI}, we adopt numerical approaches. An advantage of investigating the Bayesian version is that the variable to be estimated, namely the transition frequency $\omega_{0}$, is integrated out so that both the classical and the quantum Fisher Information are independent of $\omega_{0}$. As one can observe in Eqs. \eqref{BayesCFI} and \eqref{BayesDifference}, the CFI $I_{\omega_{0}}$ of the prior PDF is needed for the computation of $\tilde{\mathcal{F}}_{\omega_{0}}^{\text{Bayes}}$, but not for $\tilde{\mathcal{C}}_{\omega_{0}}^{\text{Bayes}}$. As an example,  we use the uniform
prior distribution, which implies $I_{\omega_0}=0$. Fig. \ref{fig:Bayesian} shows the results for the behaviours of $\tilde{\mathcal{F}}_{\omega_{0}}^{\text{Bayes}}$ and $\tilde{\mathcal{C}}_{\omega_{0}}^{\text{Bayes}}$ corresponding to the case, in which the magnetic field rotates in the $x$-$y$ plane. Both quantities exhibit close resemblance to the frequentist counterparts shown in Fig. \ref{fig:FI}, where one can observe shifts in positions of maxima of $\tilde{\mathcal{F}}_{\omega_{0}}^{\text{freq}}$ and $\tilde{\mathcal{C}}_{\omega_{0}}^{\text{freq}}$. A key observation is that the regions of large values of the Bayesian CFI are not separated by clear borders as in the frequentist counterpart shown in Fig. \ref{fig:FI}.  In a similar manner to the frequentist version, $\tilde{\mathcal{F}}_{\omega_{0}}^{\text{Bayes}}$ is usually very small in regions where $\tilde{\mathcal{C}}_{\omega_{0}}^{\text{Bayes}}$ attains small values. Despite the similarity in the pattern, one can still identify certain areas where $\tilde{\mathcal{F}}_{\omega_{0}}^{\text{Bayes}}$ attains large value while $\tilde{\mathcal{C}}_{\omega_{0}}^{\text{Bayes}}$ is kept small due to the shifts in the pattern. In the case of the other priors, 
$I_{\omega_0} \neq 0$, but it is still independent of the control parameters, and $\tilde{\mathcal{F}}_{\omega_{0}}^{\text{freq}}$ is similarly blurred as Fig. \ref{fig:Bayesian}  by integrating over the Gaussian and Jeffrey's prior. To optimize these bounds, the Bayesian version of Eq. \eqref{eq:costalphafreq} can be studied. However, these bounds are rarely saturated, and therefore, in this work, we omit this. 

{\it Minimum Mean-Square Error Estimator.} First, we study the effects of different priors on the MMSE estimator $\hat{\omega}_{0,\text{MMSE}}({\bf x})$. Fig. \ref{fig:MSSE} shows the estimates of $\tilde{\omega}_{0}$ with respect to $\bar{x}$. Only $N=8$ data points are collected in the computation of the MMSE estimates so that the effects of the prior are not suppressed by the PMF of the data. Since the support of the parameter space of the priors is $\mathbb{R}^{+}$, it is necessary to introduce cutoffs to enable numerical integration to be performed. The lower and upper limits for the Gaussian and the Jeffrey's prior are chosen to be the same as the support of the uniform prior to allow a fair comparison between the different priors. In particular, the MMSE estimate for the Gaussian prior has its intercept of the vertical axis being close to the mean of the Gaussian prior. This is due to the following facts: $p(1\vert\omega_{0})$ has a maximum at $\omega=\omega_0$, see Eq. \eqref{C0}, and for large values of $\omega_0$ it is approximately equal to zero; the posterior PDF $p(\omega_0 \vert {\bf x})$ for $\bar{x}=0$ and $\omega_{\text{\tiny{mean}}}>\omega$ is dominated by the Gaussian prior, because $\left[1-p(1\vert\omega_{0})\right]^N \approx 1$ for $\omega_0 \in [\omega_{\text{\tiny{mean}}}-2\sigma,\omega_{\text{\tiny{mean}}}+2\sigma]$. At $\bar{x}=1$, $p(\omega_0 \vert {\bf x})$ is dominated by $p(1\vert\omega_{0})^N$ and slightly distorted by the tail of the Gaussian prior and therefore the MMSE estimate gets closer to $\omega$. In comparison, the behavior of the MMSE estimators for the uniform prior and Jeffrey's prior resemble each other. They both have horizontal asymptotes close to the angular speed $\omega$, where the posterior PDF $p(\omega_0 \vert {\bf x})$ is again dominated by $p(1\vert\omega_{0})^N$.

Finally, we calculate the Bayesian mean-square error (BMSE), which reads
\begin{eqnarray}
    &&E\left[ \big(\hat{\omega}_{0,\text{MMSE}}({\bf x}) - \omega_0\big)^2 \right] \label{eq:BMSE}\\
&&=\int \left[\hat{\omega}_{0,\text{MMSE}}({\bf x}) - \omega_0\right]^2 p(\omega_0 \vert {\bf x}) p({\bf x})\,d \omega_0 \,d {\bf x} \nonumber \\
&&=\int  p({\bf x}) \,d {\bf x} \left[\int_\Omega \omega^2_0 \,p(\omega_0 \vert {\bf x}) \,d \omega_0 -  \hat{\omega}^2_{0,\text{MMSE}}({\bf x})  \right], \nonumber
\end{eqnarray}
where
\begin{equation}
 p({\bf x})= \int_\Omega p( {\bf x}\vert \omega_0) p(\omega_0) \,d \omega_0.   
\end{equation}
The meaning of the notation $\int  p({\bf x}) \,d {\bf x}$ in \eqref{eq:BMSE} based on Eq. \eqref{FrequentistLikelihood} is 
\begin{equation}
 \int  p({\bf x}) \,d {\bf x}=\sum^N_{k=0} p\left(k=\sum^N_{i=1} x_i\right)=1.  
\end{equation}

\begin{figure}[t!]
    \includegraphics[width=0.48\textwidth]{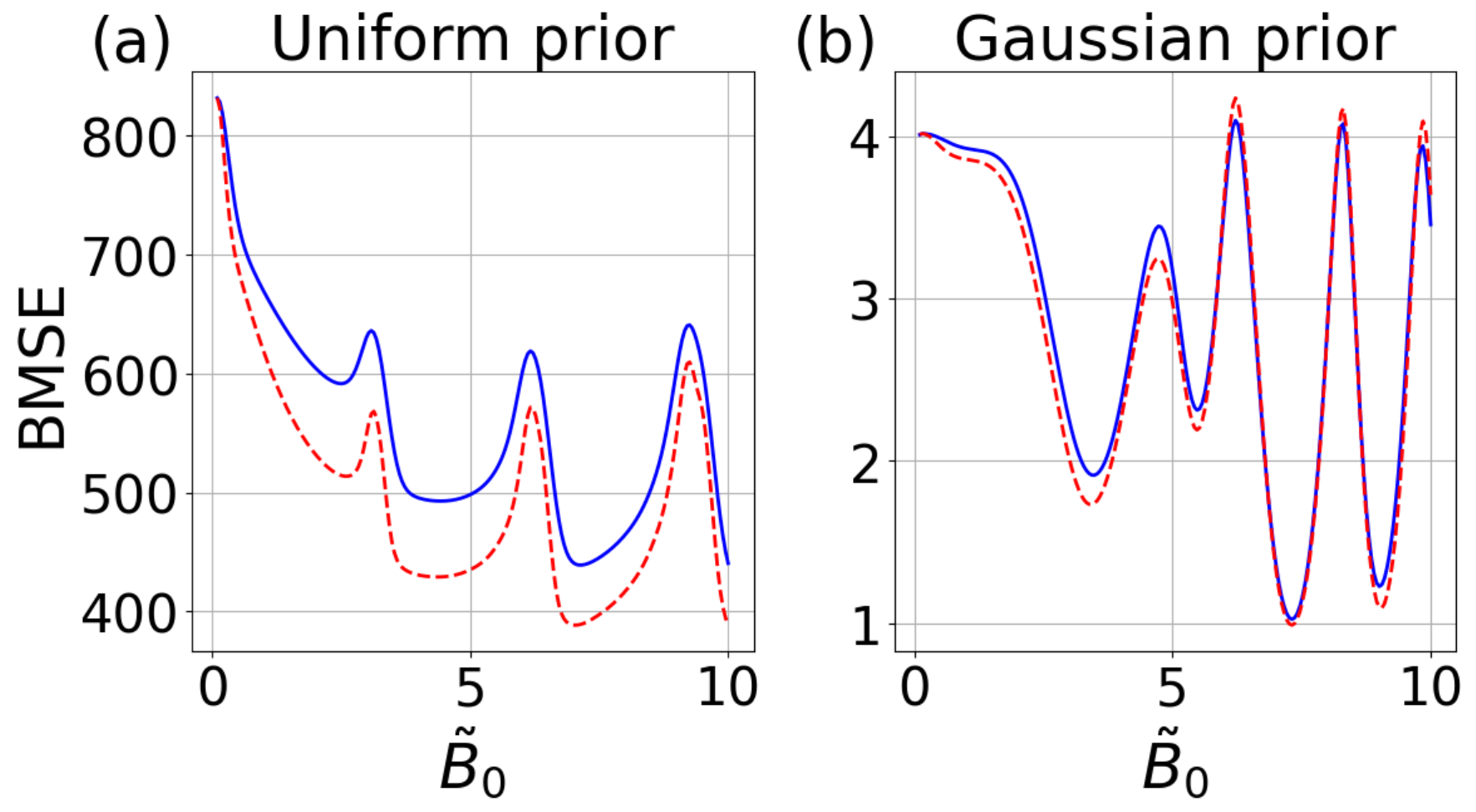}
    \caption{Bayesian mean-square error (BMSE) of the MMSE estimator as a function of the magnetic-field strength $\tilde{B}_{0}$ for (a) the uniform prior and (b) the Gaussian prior. The blue solid curve and the red dashed curve correspond to $N=15$ and $N=40$ data points respectively. The values of the other parameters are identical to those in Fig. \ref{fig:MSSE}. It can be observed that the BMSE exhibits oscillatory variations with respect to an increase in $\tilde{B}_{0}$. The reduction in the BMSE due to an increase in the number of the data points is less significant at certain values of $\tilde{B}_{0}$ . In the case of the Gaussian prior, an increase in $N$ results in a noticeable improvement in the BMSE only when $\tilde{B}_{0}$ is small. For large values of $\tilde{B}_{0}$, the sample size has little impact on the curves, and they are essentially identical, apart from minor numerical inaccuracies.}
    \label{fig:BMSE}
\end{figure}

Results of the BMSE for the uniform and the Gaussian prior concerning the variation of the magnetic-field strength $\tilde{B}_{0}$ for the cases of 15 and 40 data points are presented in Fig. \ref{fig:BMSE}. The increase in the number of data points results in an observable reduction in the BMSE as anticipated. It is interesting to note that the informative Gaussian prior improves the MMSE estimator, i.e., any prior knowledge in the Bayesian sense reduces the BMSE \cite{Kay}.

{\it Maximum a Posteriori Estimator.} As described in Sec. \ref{4}, the MAP estimator is obtained upon the maximization of the posterior PDF and thus its derivative with respect to $\omega_0$ is calculated, which is given by
\begin{eqnarray}\label{MAP}
    \frac{\partial \ln p(\omega_{0} \vert \mathbf{x})}{\partial \omega_{0}} &=& \frac{k-Np(1\vert\omega_{0})}{p(1\vert\omega_{0})\left[1-p(1\vert\omega_{0})\right]} \frac{\partial p(1\vert\omega_{0})}{\partial \omega_{0}} \nonumber \\
    &+& \frac{\partial \ln p(\omega_{0})}{\partial \omega_{0}},
\end{eqnarray}
where $p(1\vert\omega_{0}) \neq 0$ or $1$. Demanding this derivative to vanish at the MAP estimate, i.e.
\begin{equation}
    \left. \frac{\partial \ln p(\omega_{0} \vert \mathbf{x})}{\partial \omega_{0}} \right|_{\omega_0=\hat {\omega}_{0,\text{\tiny{MAP}}}({\bf x})} = 0,
\end{equation}
we obtain
\begin{eqnarray}
   &&\left. - \frac{1}{N} \frac{\partial \ln p(\omega_{0})}{\partial \omega_{0}} \frac{p(1\vert\omega_{0})\left[1-p(1\vert\omega_{0})\right]}{\frac{\partial p(1\vert\omega_{0})}{\partial \omega_{0}}} \right|_{\omega_0=\hat {\omega}_{0,\text{\tiny{MAP}}}({\bf x})} \nonumber \\
   &&+\left.  p(1\vert\omega_{0}) \right|_{\omega_0=\hat {\omega}_{0,\text{\tiny{MAP}}}({\bf x})} =\bar{x},
\end{eqnarray}
where we assume that
\begin{equation}
\left. \frac{\partial p(1\vert\omega_{0})}{\partial \omega_{0}} \right|_{\omega_0=\hat {\omega}_{0,\text{\tiny{MAP}}}({\bf x})}\neq 0,  
\end{equation}
because otherwise the MAP estimate does not depend on the data. In the asymptotic limit $N \to \infty$ or when the derivative $\frac{\partial \ln p(\omega_{0})}{\partial \omega_{0}} = 0$, one recovers the ML estimator.
This is a consequence of the Bernstein–von Mises theorem \cite{Vaart}, which shows that the posterior PDF converges in total variation distance as $N\to \infty$ to a Gaussian distribution centered at the ML estimator with a variance equal to the Cram\'er-Rao bound. This result shows that for an infinitely large dataset the prior is completely suppressed by the PMF of the data. In the case of the uniform prior, the derivative $\frac{\partial \ln p_{u}(\omega_{0})}{\partial \omega_{0}} = 0$ vanishes. Thus, the MAP estimator for a uniform prior is equal to the ML estimator and satisfies the second derivative test of Eq. \eqref{eq:2ndtest}.

\begin{figure}[t!]
    \includegraphics[width=0.48\textwidth]{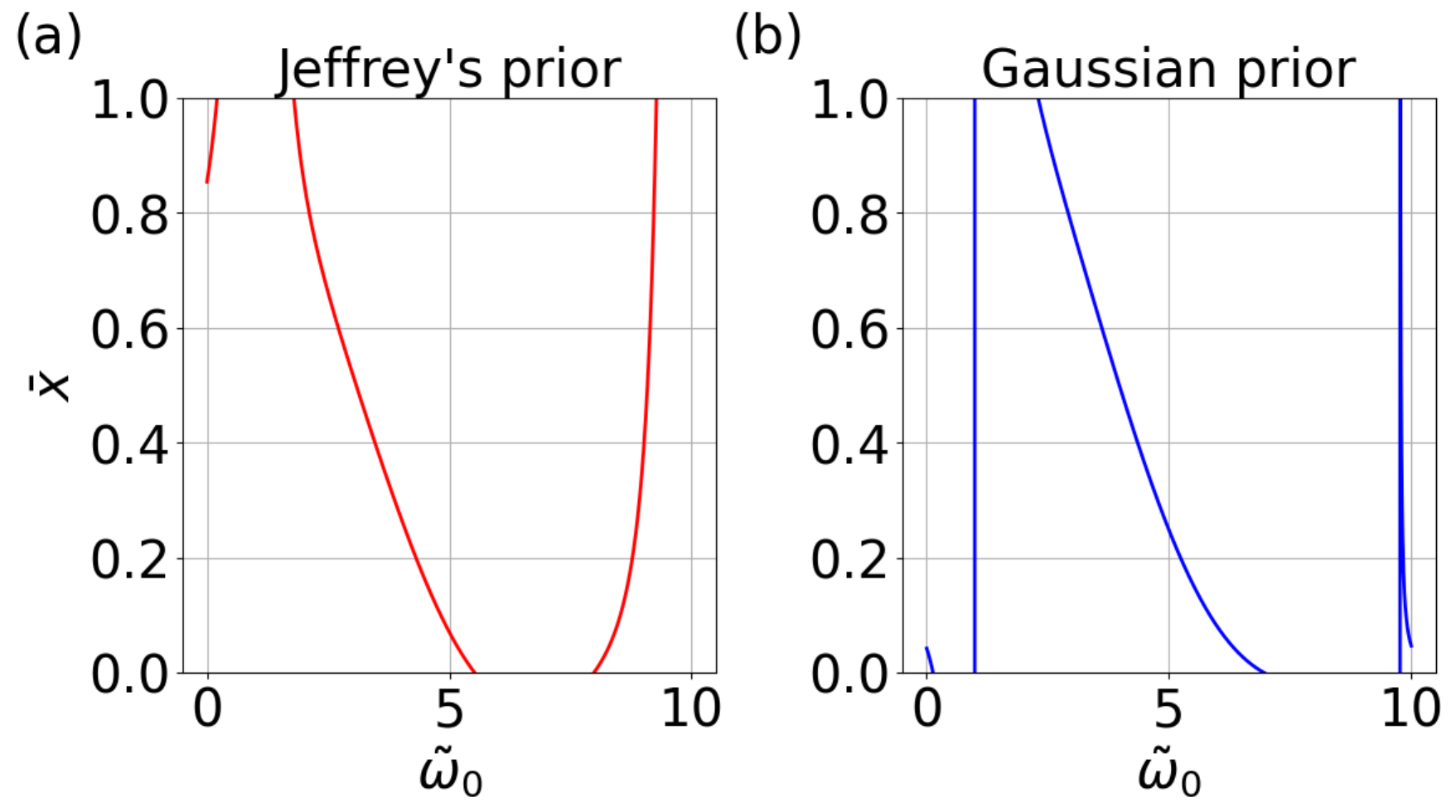}
    \caption{The variation of the average photon count rate $\bar{x}$ with respect to $\tilde{\omega}_{0}$ for (a) the Jeffrey's prior in Eq. \eqref{eq:JeffMAP} and (b) the Gaussian prior in Eq. \eqref{eq:GaussMAP}. Given the average photon count rate $\bar{x}$ determined by the measurements, one constructs the corresponding horizontal line which crosses the curves shown in the figures. The $\tilde{\omega}_{0}$-coordinates of the intersections correspond to the MAP estimates.
    The values of the parameters are chosen to be the same as those in Fig. \ref{fig:MSSE}.}
    \label{fig:MAP}
\end{figure}

We then consider the Jeffrey's prior. Computing the derivative of the logarithm of the posterior PDF given in Eq. \eqref{posterior} and setting it to zero gives
\begin{eqnarray}\label{eq:JeffMAP}
&&\left. - \frac{1}{N} \frac{p(1\vert\omega_{0})\left[1-p(1\vert\omega_{0})\right]}{\frac{\partial p(1\vert\omega_{0})}{\partial \omega_{0}}} \frac{\partial \ln \left \vert \frac{\partial p(1\vert\omega_{0})}{\partial \omega_{0}} \right \vert}{\partial \omega_{0}}\right|_{\omega_0=\hat {\omega}_{0,\text{\tiny{MAP}}}({\bf x})} \nonumber \\
   &&+\left. \left[\frac{1-2p(1\vert\omega_{0})}{2N} +p(1\vert\omega_{0}) \right]\right|_{\omega_0=\hat {\omega}_{0,\text{\tiny{MAP}}}({\bf x})} =\bar{x}.
\end{eqnarray}

In the case of the Gaussian prior, repeating the above procedure yields
\begin{eqnarray}\label{eq:GaussMAP}
&& \Bigg\{ \frac{\omega_{0}-\omega_{\text{\tiny{mean}}}}{N\sigma^{2}}\frac{p(1\vert\omega_{0})\left[1-p(1\vert\omega_{0})\right]}{\frac{\partial p(1\vert\omega_{0})}{\partial \omega_{0}}} \nonumber \\
&& \qquad +p(1\vert\omega_{0})\Bigg\}\Bigg|_{\omega_0=\hat {\omega}_{0,\text{\tiny{MAP}}}({\bf x})} =\bar{x}. 
\end{eqnarray}

The above expressions for $\bar{x}$ are plotted against $\tilde{\omega}_{0}$ and the results are shown in Fig. \ref{fig:MAP}, in which the parameter values used are identical to those in Fig. \ref{fig:MSSE}. One can observe that a given value of $\bar{x}$ leads to multiple estimates for $\tilde{\omega}_{0}$ with the use of the MAP estimator. Extra knowledge of the TLS is required to rule out inappropriate estimates. Furthermore, readers should be reminded to check if the second-order derivative of the posterior PDF with respect to the estimated parameter is strictly negative, since it is possible that the solutions for Eqs. \eqref{eq:JeffMAP} and \eqref{eq:GaussMAP} correspond to a local minimum of the posterior PDF. In case the second-order derivative vanishes, it is inconclusive if the local maximum is attained. 

In the numerical investigations presented in this section, only dimensionless quantities are deployed. Readers are reminded that time dependence is manifested in these dimensionless quantities. The time of the operating gate window of the photon detector at which the measurement data are collected can lead to an observable impact on the accuracy, since this time has influence on all the results through the dimensionless control variables. To obtain the estimate of the transition frequency $\omega_0$, the dimensionless estimates have to be divided by the operating time of the photon detector. \\

\section{Conclusion}\label{5}

In this article, we have investigated the estimation of the transition frequency of a two-level quantum system (TLS) from the perspective of Bayesian and frequentist approaches. The prototype model consists of a TLS being placed in a magnetic field, which rotates at a constant angular speed at a fixed angle about the axis of gyration. These parameters are steerable and we have used this controllability of the system to make our estimation strategies optimal. The emitted photons are collected by a single-photon detector and this results in a binary output data of the system. The probability mass function of this data carries the information
about the unknown transition frequency and the estimators are functions of the dataset. 

The Cram\'er-Rao lower bound is an important benchmark in estimation theory and therefore we have studied the deviation between the quantum (QFI) and the classical Fisher information (CFI). The quantum measurement scenario is fixed, because we detect only the emitted photons, i.e., the probability of the TLS being either in the ground or excited state. The CFI is a measure of the accuracy in the classical postprocessing of the data, while the QFI is its upper bound.  It has been found that the maxima of the CFI are often attained at places at which the above-mentioned deviation admits large values. This can lead to a conflict in the choice of the optimal strategy. The worst-case scenario is a situation, where the deviation between the two quantities is very small, which seems to imply an optimal situation, but the value of the CFI is also small. This conflict of interest can be easily overlooked. Therefore, we have studied a way of how to keep the CFI large and the deviation small by introducing a weight parameter $\alpha$, see Eq. \eqref{eq:costalphafreq}. In the extreme case, when $\alpha$ is equal to zero (one), the cost function is solely determined by the reciprocal of the CFI (the deviation). In the frequentist approach, we have presented some optimal values of the control parameters depending on $\alpha$. 

We then considered several typical frequentist and Bayesian estimators. One of the challenges is that the probability mass function of the data is a nonlinear function of the transition frequency. This has caused several issues. We have shown that the minimum variance unbiased estimator based on the derivation of the Cram\'er-Rao lower bound cannot be found. However, the maximum likelihood (ML) estimator exists, which is asymptotically consistent and efficient, but a sufficiently large set of photon counts may be a technical challenge. Unfortunately, the ML estimator does not always deliver unambiguous estimates, but we have presented a possible solution with the help of the control parameters.  We then had a look at the Bayesian approach, namely, the minimum mean-square error (MMSE) and the maximum {\it a posteriori} (MAP) estimators. The MAP estimator, which is the counterpart of the ML estimator in Bayesian statistics, suffers also from the same ambiguity issue. We have shown this for three different prior probability distribution functions of the transition frequency, namely uniform, Jeffery's, and Gaussian distribution functions. In contrast, the MMSE estimator is known to be optimal for the Bayesian mean-square error and does not suffer from the problem of multiple estimates. It is demonstrated for the noninformative priors, i.e., the uniform and Jeffrey's prior, that the estimates have very similar properties as a function of the measured data. In case the average photon count rate is equal to one, the numerical results suggest that the estimate of the transition frequency of the TLS approximately matches the frequency of the gyrating magnetic field. The Gaussian distribution function is an informative prior and imposes constraints on the estimates, which appear at zero photon count to be approximately equal to the mean of the prior. We have also studied the Bayesian mean-square error of the MMSE estimator for different number of data points. During this combined numerical and analytical investigations, we have used dimensionless parameters, and thus the time of a single-photon detection, i.e., the detector's reset time, has to be used to compute the value of the transition frequency. 

Given the approach considered here, we conclude that a nonheuristic approach to constructing efficient estimators is challenging even for quantum systems with an analytically solvable model. This is because the probability mass function of the data is a nonlinear function of the unknown parameter. However, the MMSE estimator of Bayesian statistics is always employable, efficient for enough data points, and delivers unambiguous estimates, but the inferences about the unknown {\it random} parameter are made from its posterior distribution. This suggests that the Bayesian approach has some technical advantage over the frequentist one in such cases and thus investigations for the search of proper priors ought to be considered in future applications. The practical use of the discussed estimators in experimental quantum physics largely depends on the technical challenges encountered in the implementation. While finding the MVU estimator can be challenging, it and the MMSE estimator are commonly used in signal processing \cite{Biguesh, Katselis, Shoari}.  On the other hand, the ML estimator is widely applied in quantum tomography \cite{Badurek2004}, quantum metrology \cite{Vidrighin, Yin}, and quantum sensing \cite{Gefen}. The MAP estimator is generally used less frequently, although there are experimental instances to which it is applied \cite{Gursoy, Chan}. For TLS, such as in the fabrication and characterization of qubits \cite{Zwerver,VanDamme}, the heuristic least squares method is often employed. A key feature of this method is that it does not rely on probabilistic distribution of the data, but  assumes only the functions to be fitted to the measurement results. Possible applications of our approach to the characterization of qubits require a statistical model for the data.  

Finally, it is important to stress that only a semi-classical treatment of the model is employed and the effects of measurement backaction are not considered in this study. In the context of the semiclassical approach, the gyrating magnetic field is modeled as a classical field so that only stimulated emission of photons can take place. Should a full quantum treatment be considered, the TLS can emit photons via spontaneous emission.  Furthermore, the state of the TLS will depend strongly on the state of the electromagnetic field. Continuous observation of a quantum system exerts a backaction \cite{Mensky, Barchielli, Diosi, Belavkin, Presilla, Jacobs, Bernad}, which, for example, inhibits the transition in the TLS \cite{Presilla} or causes an experimentally measurable frequency shift \cite{Bushev}. Parameter estimation in an extended quantum model is another topic for future investigation.

\begin{acknowledgments}
This work was supported by AIDAS-AI, Data Analytics
and Scalable Simulation, which is a Joint Virtual Laboratory
gathering the Forschungszentrum J\"ulich and the French
Alternative Energies and Atomic Energy Commission, by the Ministry of Culture and Innovation, and the
National Research, Development and Innovation Office
within the Quantum Information National Laboratory
of Hungary (Grant No. 2022-2.1.1-NL-2022-00004), as well as by the German Federal Ministry of Research (BMBF) under the project SPINNING (No. 13N16210). We are grateful to A. B. Frigyik, C. M. Sanavio, F. Preti, and R. Pal for stimulating discussions.

\end{acknowledgments}

\appendix

\section{Time evolution of the TLS}
\label{AppendixA}

In this Appendix, we present the well-known approach and solution to the Schr\"odinger equation of Sec. \ref{2}. We expand the state vector as $\ket{\Psi(t)} = c_{0}(t)\ket{0} + c_{1}(t) \ket{1}$, and thus the Sch\"{o}dinger equation can be written as a system of two coupled ordinary differential equations (ODEs): 
\begin{equation}\label{SE1}
    \frac{dc_{0}(t)}{dt} = -i a c_{0}(t) - i b e^{-i\omega t} c_{1}(t) 
\end{equation}
and
\begin{equation}\label{SE2}
     \frac{dc_{1}(t)}{dt} = iac_{1}(t) - ibe^{i\omega t} c_{0}(t),
\end{equation}
where $a = \omega_{0}/2 + \gamma B_{0}\cos \theta $ and $b =\gamma B_{0} \sin \theta$. Differentiating \eqref{SE1} with respect to time, making use of \eqref{SE1} and \eqref{SE2} and upon rearrangement leads to the second order ODE
\begin{equation}\label{2ndc0}
    \frac{d^{2}c_{0}(t)}{dt^{2}} + i \omega \frac{dc_{0}(t)}{dt} + [a^{2} + b^{2} - \omega a] c_{0}(t) = 0 .
\end{equation}

The general solution of \eqref{2ndc0} is then
\begin{equation}\label{GS}
     c_{0}(t) = e^{-\frac{i \omega t}{2}} (A e^{\frac{iqt}{2}}+ Be^{-\frac{iqt}{2}}),
\end{equation}
where $q = \sqrt{(\omega_{0}- \omega)^{2} + 4 \gamma B_{0} \cos \theta (\omega_{0}- \omega) + 4 \gamma^{2} B_{0}^{2}}$. Applying the Euler formula $e^{ix} = \cos x + i \sin x$ to \eqref{GS} and defining $A'= A + B$ and $B'= i(A - B)$, we obtain 
\begin{equation}\label{c0}
    c_{0}(t) = e^{-\frac{i \omega t}{2}} \left[ A' \cos \left(\frac{qt}{2} \right) + B' \sin \left(\frac{qt}{2} \right) \right].
\end{equation}
We substitute \eqref{c0} into \eqref{SE1} to get
\begin{eqnarray}
    &&c_{1}(t) = \frac{i e^{\frac{i\omega t}{2}}}{\gamma B_{0} \sin \theta}  \\
    && \times \left\{ \left[ i \left( \frac{\omega_{0} - \omega}{2} + \gamma B_{0} \cos \theta \right) A' + \frac{q}{2} B' \right] \cos \left( \frac{qt}{2} \right) \right. \nonumber \\
    && + \left. \left[ i \left( \frac{\omega_{0} - \omega}{2} + \gamma B_{0} \cos \theta \right)B' - \frac{q}{2}A'  \right] \sin \left( \frac{qt}{2} \right) \right\}. \nonumber
\end{eqnarray}
Our initial condition is $\ket{\Psi(0)}=\ket{1}$, or alternatively $c_{0}(0) = 0$ and $c_{1}(0) = 1$, which leads to 
\begin{eqnarray}
    A'&=&0, \nonumber \\
    B'&=& -i\frac{2 \gamma B_{0} \sin \theta}{q}. \nonumber
\end{eqnarray}
The solution to our initial-value problem is therefore
\begin{equation}
    c_{0}(t) = -i e^{-\frac{i\omega t}{2}} \left( \frac{2 \gamma B_{0} \sin \theta}{q} \right) \sin \left( \frac{qt}{2} \right)
\end{equation}
and
\begin{eqnarray}
     c_{1}(t) &=& e^{\frac{i\omega t}{2}} \left[ \cos \left(\frac{qt}{2} \right) \right. \\
    &-& \left. i  \left( \frac{\omega-\omega_{0} - 2 \gamma B_{0} \cos \theta}{q}\right) \sin \left(\frac{qt}{2} \right) \right]. \nonumber
\end{eqnarray}
The density matrix $\rho = \ket{\Psi} \bra{\Psi}$ can then be expressed in terms of the components of the state vector as
\begin{align}
    \rho = 
    \begin{pmatrix}
        \rho_{00}  &
        \rho_{01}  \\
        \rho_{01}^{*} & 
        1 - \rho_{00}
     \end{pmatrix} &=
    \begin{pmatrix}
        \vert c_{0} \vert^{2}  &
        c_{0} c_{1}^{*}  \\
        c_{0}^{*}c_{1} & 
        \vert c_{1} \vert^{2}
    \end{pmatrix} \notag \\
    &= 
    \begin{pmatrix}
        \vert c_{0} \vert^{2}  &
        c_{0} c_{1}^{*}  \\
        (c_{0} c_{1}^{*})^{*} & 
        1 - \vert c_{0} \vert^{2}
     \end{pmatrix} .
\end{align}
This allows us to identify $\rho_{00}$ and $\rho_{01}$ as 
\begin{equation}
    \rho_{00} = \vert c_{0} \vert^{2} = \frac{4\gamma^{2}B_{0}^{2} \sin^{2} \theta}{q^{2}} \sin^{2} \left( \frac{qt}{2} \right)
\end{equation}
and 
\begin{eqnarray}
    &&\rho_{01} = c_{0} c_{1}^{*} = \frac{\gamma B_{0} \sin \theta}{q} e^{-i\omega t}  \\
    && \times \left\{\left(\frac{\omega - \omega_{0} - 2 \gamma B_{0} \cos \theta}{q}\right) \left[1- \cos (qt)\right] -i \sin (qt) \right\}. \nonumber
\end{eqnarray}

\section{Regularity Condition}
\label{AppendixB}

We need to show that the regularity condition is satisfied by the joint
PMF $ p({\bf x}; \omega_0)$ of the data  to get Eq. \eqref{eq:likelihood,omega}. Our starting point is to rewrite the PMF as
\begin{eqnarray}
\label{eq:C1}
    p(\mathbf{x}; P_{1})= \binom{N}{k} P^k_1 (1-P_1)^{N-k}
\end{eqnarray}
where the total photon count $k=\sum_{i=1}^{N} x_{i}$ occurs during $N$ trials and $P_{1} = p(1;\omega_{0}) = \vert c_{0} \vert^{2}$. Note that
\begin{widetext}
\begin{eqnarray}
     \mathbb{E}\left[\frac{\partial \ln p(\mathbf{x}; P_{1})}{\partial P_{1}}\right] &=& \sum_{x_{1}=0}^{1} \cdots \sum_{x_{N}=0}^{1} p(\mathbf{x}; P_{1}) \frac{\partial \ln p(\mathbf{x}; P_{1})}{\partial P_{1}}  = \sum_{x_{1}=0}^{1} \cdots \sum_{x_{N}=0}^{1} p(\mathbf{x}; P_{1}) \cdot\left[\frac{k}{P_{1}} - \frac{N-k}{1-P_{1}} \right] \nonumber \\
    &=& \sum_{x_{1}=0}^{1} \cdots \sum_{x_{N}=0}^{1} p(\mathbf{x}; P_{1}) \cdot\left[ \left(\frac{1}{P_{1}} + \frac{1}{1-P_{1}} \right) k - \frac{N}{1-P_{1}} \right] \nonumber \\
    &=& \sum_{x_{1}=0}^{1} \cdots \sum_{x_{N}=0}^{1} p(\mathbf{x}; P_{1}) \cdot\left[ \frac{1}{P_{1}(1-P_{1})}  \sum_{i=1}^{N} x_{i} - \frac{N}{1-P_{1}} \right]  ,
\end{eqnarray}
where $\mathbb{E}[\cdot]$ denotes the expectation taken respect to $p(\mathbf{x};P_{1})$.
In the case of i.i.d. observations and based on Eq. \eqref{eq:C1}, we can separate the PMF into a product
\begin{equation}
    p(\mathbf{x};P_{1}) \propto \prod_{j=1}^{N} p(x_{j}; P_{1}) .
\end{equation}
Then, we have
\begin{eqnarray}
    &&\sum_{x_{1}=0}^{1} \cdots \sum_{x_{N}=0}^{1} p(x_{1};P_{1}) \cdots p(x_{N};P_{1}) \cdot \left[ \frac{1}{P_{1}(1-P_{1})}  \sum_{i=1}^{N} x_{i} - \frac{N}{1-P_{1}} \right] \\
    &&= \sum_{x_{1}=0}^{1} \cdots \sum_{x_{N}=0}^{1}  \left[ \frac{1}{P_{1}(1-P_{1})}  \cdot \sum_{i=1}^{N} x_{i} p(x_{1};P_{1}) \cdots p(x_{N};P_{1})  - \frac{N}{1-P_{1}} p(x_{1};P_{1}) \cdots p(x_{N};P_{1}) \right] \nonumber \\
    &&= \frac{1}{P_{1}(1-P_{1})} \cdot \sum_{i=1}^{N} \sum_{x_{i}=0}^{1} x_{i} p(x_{i};P_{1})   - \frac{N}{1-P_{1}} \sum_{x_{1}=0}^{1} p(x_{1};P_{1}) \cdots \sum_{x_{N}=0}^{1} p(x_{N};P_{1}) \nonumber \\ 
    &&= \frac{1}{P_{1}(1-P_{1})} \sum_{i=1}^{N} P_{1} - \frac{N}{1-P_{1}} = 0,
\end{eqnarray}
where we have exchanged the summation signs and used the relations
\begin{equation}
    \sum_{x_{i}=0}^{1} p (x_{i}; P_{1}) = 1  \quad \text{and} \quad
    \sum_{x_{i}=0}^{1} x_{i} p(x_{i}; P_{1}) = P_{1}.
\end{equation}
\end{widetext}
We arrive then at the regularity condition
\begin{equation}
 \mathbb{E}\left[\frac{\partial \ln p(\mathbf{x}; P_{1})}{\partial P_{1}}\right]=0, 
\end{equation}
which implies with the help of the chain rule
\begin{equation}
 \mathbb{E}\left[\frac{\partial \ln p(\mathbf{x}; \omega_0)}{\partial \omega_0}\right]=0.
\end{equation}

The MVU estimator for $P_{1}$ is obtained by taking the derivative of $\ln p(\mathbf{x};P_{1})$ with respect to $P_{1}$ gives 
\begin{eqnarray}
    \frac{\partial}{\partial P_{1}} \ln p(\mathbf{x} ; P_1)
    = \frac{N}{P_{1}(1-P_{1})} \left[ \frac{1}{N} \sum_{i=1}^{N} x_{i} - P_{1}  \right].
\end{eqnarray}
Now, based on the equation of the MVU estimator \cite{Kay}
\begin{equation}
 \frac{\partial}{\partial P_{1}} \ln p(\mathbf{x} ; P_1)=I(P_1) \left[ \hat{P}_{1,\text{\tiny{MVU}}}({\bf x})-P_1\right],   
\end{equation}
we can see that average photon count rate $\bar{x} = \frac{1}{N}\sum_{i=1}^{N} x_{i}$ is an MVU estimator for $P_{1}$ with variance $\frac{P_{1}(1-P_{1})}{N}$.

\section{Classical and the Quantum Fisher Information}
\label{AppendixC}

We have argued in Section \ref{3} that there are two possible projective measurements. Thus, with the help of the corresponding probabilities $p(1;\omega_{0}) = \rho_{00}$ and $p(0;\omega_{0}) = 1-\rho_{00}$, the classical Fisher Information for a single measurement result yields
\begin{align}\label{CFI1}
    \mathcal{F}_{\omega_{0}}^{\text{freq}} &= p(1;\omega_{0}) \left[\frac{\partial \ln p(1;\omega_{0})}{\partial \omega_{0}} \right]^{2} + p(0;\omega_{0}) \left[\frac{ \partial \ln p(0;\omega_{0})}{\partial \omega_{0}} \right]^{2} \notag \\
    &= \frac{1}{p(1;\omega_{0})} \left[\frac{\partial p(1;\omega_{0})}{\partial \omega_{0}} \right]^{2} + \frac{1}{p(0;\omega_{0})} \left[\frac{\partial p(0;\omega_{0})}{\partial \omega_{0}} \right]^{2} \notag \\
    &= \frac{1}{\rho_{00}} \left[\frac{\partial \rho_{00}}{\partial \omega_{0}} \right]^{2} + \frac{1}{1-\rho_{00}} \left[\frac{\partial}{\partial \omega_{0}} (1-\rho_{00})\right]^{2} \notag \\
    &= \left[ \frac{1}{\rho_{00}} + \frac{1}{1-\rho_{00}}  \right] \left[\frac{\partial}{\partial \omega_{0}}\rho_{00} \right]^{2} \notag \\
    &= \frac{1}{\rho_{00}(1-\rho_{00})} \left[\frac{\partial \rho_{00}}{\partial \omega_{0}} \right]^{2}.  
\end{align}

In the case of the estimation of the transition frequency $\omega_{0}$, the QFI reads
\begin{equation}\label{QFI_1pa}
    \mathcal{H}_{\omega_{0}}^{\text{freq}} = \Tr [L_{\omega_{0}}^{2} \rho],
\end{equation}
where
\begin{equation}  
\label{DenMatDer1}
    \frac{d \rho}{d \omega_0} = \frac{1}{2} \{\rho, L_{\omega_0}\}
\end{equation}
and $\{,\}$ stands for the anticommutator. A pure state in the density-matrix form has the idempotent property $\rho = \rho^{2}$. Now, differentiating both sides with respect to $\omega_{0}$ results in
\begin{align}\label{DenMatDer2}
    \frac{\partial \rho}{\partial \omega_{0}} &= \rho \frac{\partial \rho}{\partial \omega_{0}} + \frac{\partial \rho}{\partial \omega_{0}} \rho \notag \\
    &= \frac{1}{2} \left[\rho \left(2\frac{\partial \rho}{\partial \omega_{0}} \right) + \left( 2 \frac{\partial \rho}{\partial \omega_{0}}\right) \rho \right] \notag \\
    &= \frac{1}{2}\left\{\rho, 2 \frac{\partial \rho}{\partial \omega_{0}} \right\} .
\end{align}
A direct comparison of \eqref{DenMatDer1} and \eqref{DenMatDer2} allows us to identify and compute the SLD operator as 
\begin{align}
     L_{\omega_{0}} = 2 \frac{\partial \rho}{\partial \omega_{0}} 
    &=  2 \frac{\partial }{\partial \omega_{0}}\begin{pmatrix}
        \rho_{00} & \rho_{01} \\
        \rho_{01}^{*} & 1 - \rho_{00}
    \end{pmatrix}\\
    &=2
    \begin{pmatrix}
        \frac{\partial \rho_{00}}{\partial \omega_{0}}  &
        \frac{\partial \rho_{01}}{\partial \omega_{0}}  \\
        \frac{\partial \rho_{01}^{*}}{\partial \omega_{0}} & 
        - \frac{\partial \rho_{00}}{\partial \omega_{0}} 
    \end{pmatrix}.
\end{align}
Furthermore, we have
\begin{equation}
    L_{\omega_{0}}^{2} = 4 \left( \left[\frac{\partial \rho_{00}}{\partial \omega_{0}} \right]^{2} + \left \vert \frac{\partial \rho_{01}}{\partial \omega_{0}} \right \vert^{2} \right) I_2.
\end{equation}
where $I_2$ is the $2\times 2$ identity matrix. Substituting this result into \eqref{QFI_1pa} gives us the expression for the QFI
\begin{align}\label{QFI1_1pa2}
    \mathcal{H}_{\omega_{0}}^{\text{freq}} &= \Tr \{L_{\omega_0}^{2} \rho \} \notag \\
    &= 4 \left( \left[\frac{\partial \rho_{00}}{\partial \omega_{0}} \right]^{2} + \left \vert \frac{\partial \rho_{01}}{\partial \omega_{0}} \right \vert^{2} \right) \Tr \rho \notag \\
    &= 4 \left( \left[\frac{\partial \rho_{00}}{\partial \omega_{0}} \right]^{2} + \left \vert \frac{\partial \rho_{01}}{\partial \omega_{0}} \right \vert^{2} \right),
\end{align}
where the unit trace property $\Tr \rho = 1$ is used in the second line.

\bibliography{manuscript}
\end{document}